\documentclass[12pt]{article}
\usepackage{amsmath}
\usepackage{graphicx}
\usepackage{times}
\usepackage[T1]{fontenc}
\usepackage[utf8]{inputenc}
\usepackage[margin=1cm]{caption}

\usepackage[square,numbers,sort&compress]{natbib}

\author
{Kacho Imtiyaz Ali Khan$^{1,\ast}$,  Tauqir Shinwari$^{1}$, Soheil Ershadrad$^{2}$, Majid Ahmadi$^{3}$, \\ Weiben Li$^{5}$, Hua Lv$^{1}$, Frans Munnik$^{4}$, Adriana I. Figueroa$^{6,7}$, Manuel Valvidares$^{5}$, \\ Sandra Ruiz-G\'{o}mez$^{5}$, Lucia Aballe$^{5}$, Jens Herfort$^{1}$, Michael Hanke$^{1}$, \\ Bart Kooi$^{3}$,  Biplab Sanyal$^{2}$ and Jo\~{a}o Marcelo J. Lopes$^{1,\ast}$\\
\\
\normalsize{$^{1}$Paul-Drude-Institut für Festkörperelektronik, Leibniz-Institut im Forschungsverbund} \\\normalsize{Berlin e.V, 10117, Germany}
\\
\normalsize{$^{2}$Department of Physics and Astronomy, Uppsala University} \\\normalsize{Uppsala, Sweden}
\\
\normalsize{$^{3}$Zernike Institute for Advanced Materials, University of Groningen} \\\normalsize{AG Groningen, 9747, The Netherlands}
\\
\normalsize{$^{4}$Helmholtz-Zentrum Dresden-Rossendorf, Institute of Ion Beam Physics and Materials Research} \\\normalsize{Dresden, 1328, Germany}
\\
\normalsize{$^{5}$ALBA Synchrotron Light Facility, CELLS, Cerdanyola del Vall\'{e}s,} \\\normalsize{08290, Spain}
\\
\normalsize{$^{6}$Departament de F\'{i}sica de la Mat\'{e}ria Condensada, Universitat de Barcelona,} \\\normalsize{Barcelona, 08028, Spain}
\\
\normalsize{$^{7}$Institut de Nanoci\'{e}ncia i Nanotecnologia (IN2UB), Universitat de Barcelona,} \\\normalsize{Barcelona, 08028, Spain}
\\
\normalsize{$^\ast$To whom correspondence should be addressed;} \\  
\normalsize{Emails: khan@pdi-berlin.de, lopes@pdi-berlin.de}
}

\date{}

\begin{document} 
\title{\LARGE\bfseries{Tuning Structure and Magnetism in Large-Scale 2D Ferromagnet Fe$_3$GeTe$_2$ through Ni Doping}} 

\maketitle

\begin{abstract}
Two-dimensional ferromagnets with strong perpendicular magnetic anisotropy (PMA) exhibit magnetic order down to the monolayer, beneficial for energy-efficient spintronic devices. In this work, molecular beam epitaxy has been employed to realize controlled Ni-doping in Fe$_{\text 3}$GeTe$_{\text 2}$ (FGT) films. MBE not only enables a large-scale growth of 2D films, but also allows a precise control over thickness and doping. X-ray diffraction and scanning transmission electron microscopy (STEM) reveal the formation of high-quality epitaxial films of pristine and Ni-doped FGT on graphene via van der Waals (vdW) epitaxy. Integrated differential phase contrast STEM images provide in-depth information on Ni-substitution/intercalation into the vdW-gaps. Ni incorporation in doped films results in the shrinking of both in-plane and out-of-plane lattice parameters. Superconducting Quantum Interference Device, Hall, and X-ray magnetic circular dichroism were utilized to probe the ferromagnetic properties of the films. Due to Ni-substitution/intercalation into the vdW-gaps for Ni-doped FGT films, we observed a suppression of PMA and a drastic reduction in the Curie temperature down to 50 K. Density functional theory calculations of structural/magnetic properties support the experimental observations and provide deep insights into the variations of magnetic exchange interaction parameters and atom-projected magnetocrystalline anisotropy energies upon Ni-doping.
\end{abstract}

\section*{Introduction}\label{sec1}
Two-dimensional (2D) magnetic materials and van der Waals (vdW) heterostructures have opened up new possibilities in the fields of spintronics, opto-spintronics, and quantum technologies due to the ability to manipulate spin in the 2D limit ~\cite{cabo2025roadmap,sierra_van_2021,lin_recent_2023, kurebayashi_magnetism_2022,thiel2019probing,prashant2025next_generation}. The novel magnetic and electronic properties of such 2D magnets, which can be tailored by engineering structure, strain, doping, and electric field-gating ~\cite{ngaloy2024strong, drachuck2018effect, huang2018electrical, cenker2022reversible, ghosh2024structural, deng_gate-tunable_2018, yu2022recent}, make them promising for the realization of ultra-compact and energy-efficient spin-based devices such as spin logic gates~\cite{jiang2025room, tao2020valley}, spin tunneling junctions~\cite{song2018giant, wang2018very, piquemal20172d, li2019spin}, and random-access magnetic memories~\cite{yang_two-dimensional_2022, zabel2007magnetic, gibertini_2019}. Among different 2D magnets, the ternary Fe-Ge(Ga)-Te metallic compounds are particularly attractive due to their relatively high Curie temperature ($T_{\rm C}$) and other promising features such as a large anomalous Hall effect (AHE)~\cite{kim2018large, ribeiro_large-scale_2022}, current-induced magnetization switching~\cite{algaidi2024magnetic, kajale_field-free_2024,pandey2025tunable}, and stabilization of magnetic skyrmions~\cite{liu_magnetic_2024, ding2019observation}.  

\begin{figure} [t!] 
\centering
\includegraphics[width=0.9\columnwidth]{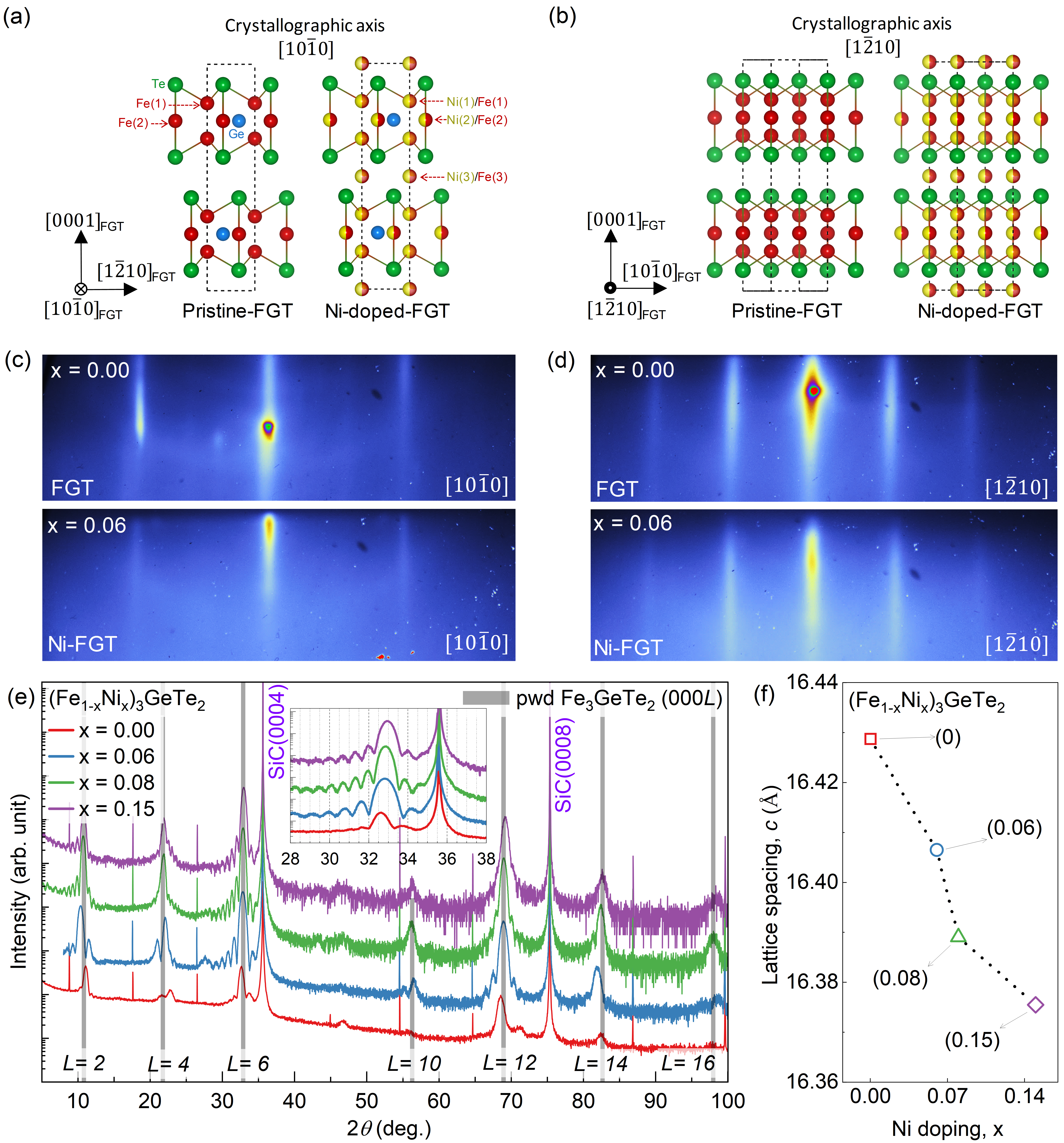}
\caption{\textbf{Schematics and structural properties of Pristine FGT and Ni-doped FGT films.} One unit cell of pristine FGT and Ni-doped FGT consists of two quintuple layers (QLs), as displayed along the two orthogonal crystallographic directions (a) $[10\overline{1}0]$ and (b) $[1\overline{2}10]$. (c,d) Corresponding reflection high-energy electron diffraction (RHEED) patterns for pristine FGT [upper panel] and Ni-doped FGT [bottom panel]. (e) Out-of-plane XRD plots measured in $\theta-2\theta$ mode, along with the calculated Bragg position for powder FGT (gray bars). The inset represents the Kiessig fringes, indicating the smooth interface and high crystalline quality of the films. (f) The variation of out-of-plane lattice constant \textit{c} as a function of Ni doping concentration in FGT films.}
\label{Fig:1}
\end{figure}

\hspace{1cm}Fe$_3$GeTe$_2$ (FGT) and Ni$_3$GeTe$_2$ (NGT) are layered materials which crystallize into the same space group $P6_3/mmc$ (No. 194)~\cite{lopes2021large, drachuck2018effect}. However, while FGT is a ferromagnetic metal with a $T_{\rm C}$ of 220~K, NGT is a paramagnetic material. The crystal structure of NGT additionally permits Ni intercalation into the vdW gaps, enabled by the partially stable interstitial Ni(3) site (Fig.~\ref{Fig:1}a,b)~\cite{deiseroth2006fe3gete2}. Interestingly, incorporation of Ni into FGT suppresses its ferromagnetic properties, including a decrease in $T_{\rm C}$ and saturation magnetization ($M_{\rm S}$), finally leading to a spin-glass state similar to the nonmagnetic phase of NGT. Although a detailed structural understanding of this magnetic evolution remains elusive, it is generally associated with Ni intercalation within the vdW gaps between FGT monolayers. This intercalation effect is highly interesting in its own right because it breaks the inversion symmetry in 2Dferromagnets, enabling an antisymmetric exchange interaction known as the Dzyaloshinskii–Moriya interaction (DMI). This interaction helps in stabilizing chiral spin textures~\cite{saha2022observation, saha2024high, zhang2024spin}, which are expected to play a pivotal role in emerging 2D spintronic technologies such as racetrack memories and neuromorphic computing. So far, the incorporation of dopants into 2D magnetic materials, which results in intercalation among other effects, has been limited to bulk single crystals and consequently micrometer-sized flakes exfoliated from them~\cite{drachuck2018effect, deiseroth2006fe3gete2, saha2022observation, saha2024high, zhang2024spin}. Besides their limited lateral sizes, which are incompatible with conventional device fabrication schemes ~\cite{wu2020neel, tan2018hard, wang_magnetic_2022}, tuning critical parameters such as thickness, doping concentration and homogeneity in such systems remains a significant challenge~\cite{cabo2025roadmap}. On the other hand, epitaxial growth methods such as molecular beam epitaxy (MBE) enable not only wafer-scale synthesis, but also provide a precise control over thickness and doping. Therefore, the use of MBE methods for the controlled large-area growth of 2D ferromagnets is highly suitable for the development of energy-efficient spintronic devices. Moreover, vdW epitaxy via MBE is highly promising for the realization of vdW heterostructures with ultra-clean and sharp interfaces combining dissimilar 2D materials ~\cite{lv_all-epitaxial_2024, lv2025proximity, shinwari2025above}. This is highly demanded as proximity-induced coupling effects should enable additional property tailoring~\cite{albarakati2019antisymmetric, zhao2023proximity, zheng_lateral_2024, pan_roomtemperature_2024, lv_all-epitaxial_2024,  alghamdi_highly_2019, lv2025proximity}.

\hspace{1cm}In this work, we use MBE-grown films to investigate the role of Ni doping for tailoring the structural and magnetic properties of [Fe$_{\text {1-x}}$Ni$_{\text x}$]$_{\text 3}$GeTe$_{\text 2}$ films over large areas. The structural characterizations confirm the formation of high-quality of both pristine and Ni-doped FGT films directly on graphene via vdW epitaxy, and also provide an in-depth insight on the structure evolution upon Ni substitution and intercalation into vdW gaps. Both experimental results [obtained from magnetization, Hall, and X-ray magnetic circular dichroism (XMCD)] and theoretical calculations show the reduction of ferromagnetic properties and its direct correlation to the structural modification via Ni incorporation. The results provide an advanced understanding on the role of controlled dopant incorporation into large-scale 2D magnetic systems. This is essential for the development of future spintronic devices with tailored properties and functionalities based on magnetic vdW heterostructures.

\section*{Results and Discussion}
\subsection*{Structural properties}
The chemical composition of pristine FGT and Ni-doped FGT was analyzed using Rutherford Backscattering Spectrometry (RBS). By fitting the RBS spectra (see Supplementary File S1), the Ni doping concentrations (x) in the [Fe$_{\text {1-x}}$Ni$_{\text x}$]$_{\text 3}$GeTe$_{\text 2}$ films were determined to be x = 0.00, 0.06, 0.08, and 0.15. Fig.~\ref{Fig:1}a-b illustrates the atomic structure of pristine FGT and Ni-doped FGT viewed along the crystallographic axes [10$\Bar{1}$0] and [1$\Bar{2}$10].  In the pristine FGT lattice, the Fe atoms occupy two inequivalent sites, labelled Fe(1) and Fe(2), forming a quintuple layer (QL) composed of Te/Fe/FeGe/Fe/Te slabs. The dotted rectangular box represents the unit cell, containing two QL of FGT. Upon Ni doping, Ni atoms replace Fe at multiple sites within the Fe sublattices, designated as Ni(1)/Fe(1), Ni(2)/Fe(2), and may also intercalate into the vdW gaps at Ni(3)/Fe(3), without disrupting the overall crystal symmetry [see the Ni-doped FGT case in Fig.~\ref{Fig:1}a]. DFT calculations were performed using the same structural configurations, considering possible Ni substitution at all Fe sublattices as well as Ni intercalation in the vdW gap, as shown in supplementary Table~S1 and Fig.~S2. The calculated formation energies of Ni substitution suggest that the outermost Fe sites have a higher probability of Ni occupation. It should also be noted that the formation energy of Ni at the vdW gap at a site aligned with outermost Fe atoms has an energy comparable to the substitutional one mentioned above. This indicates the coexistence of both substitutional  and interstitial Ni dopants within the vdW gaps, as confirmed by STEM investigations presented later. 

Reflection High-Energy Electron Diffraction (RHEED) images were acquired for both pristine FGT (upper panel) and Ni-doped FGT (bottom panel) along two orthogonal in-plane crystallographic directions, $[10\overline{1}0]$ and $[1\overline{2}10]$, as shown in Fig.~\ref{Fig:1}c and Fig.~\ref{Fig:1}d, respectively. Both films exhibit sharp, elongated RHEED streaks, confirming the atomically smooth quality of the film surface with hexagonal lattice symmetry. From the RHEED streak spacing, the in-plane lattice parameter `\textit{a}' was determined, and it was found to decrease from \textit{a} = $4.069$~\AA~  (pristine FGT, x = 0.00) to $3.931$~\AA~ (Ni-doped FGT, x = 0.15) with increasing Ni doping. The contraction in the in-plane lattice parameter as a function of Ni doping has been further verified by synchrotron-based grazing-incidence X-ray diffraction (GIXRD) measurements (see supplementary file S3). DFT structural optimizations also confirm lattice shrinkage upon Ni doping, with the in-plane lattice parameter $a$ decreasing from $4.06(3)$~\AA~ in pristine FGT to $3.98(9)$~\AA~ in NGT (x = 1.0), see supplementary Fig.~S4. In case of x = 0.15, lattice parameter of \textit{a} = $4.00(8)$~\AA~ was calculated. In addition, the presence of vacancies in Fe(Ni) sites further reduces the in-plane lattice parameter in both systems.

\hspace{1cm}Out-of-plane XRD measurements were performed in $\theta-2\theta$ mode for pristine (x $=0$) FGT and Ni-doped (x $=0.06-0.15$) FGT films grown on graphene/SiC(0001) [see Fig.~\ref{Fig:1}e]. All the samples exhibit strong Bragg reflections corresponding to FGT(000\textit{L}) ($L=$ 2, 4, 6, ..., 16) planes, confirming the highly crystalline phase. Additionally, the presence of sharp Kiessig fringes around the FGT(0002), (0006) and (00012) plane peaks indicates the formation of smooth interfaces between the films and graphene/SiC [inset Fig.~\ref{Fig:1}e]. Using Bragg's equation, the out-of-plane lattice constants `\textit{c}' were determined from the (000\textit{L}) reflections for the different samples (see Fig.~\ref{Fig:1}f). It was found that values of \textit{c} decrease systematically with increasing Ni doping. This trend is evidenced by the shift of (000\textit{L}) peaks, indicating continuous contraction along the \textit{c}-axis, similar to previous observations in bulk crystals of Ni-doped FGT~\cite{drachuck2018effect} and Ni-doped FGaT~\cite{zhu2024effect}. Similarly, DFT calculations predict a contraction along the $c$-axis upon Ni doping, which is shown in supplementary Fig.~S4. The decreasing trend agrees well with experiment, with $c$ reducing from 16.20(6)~\AA\ in pristine FGT to 15.87(7)~\AA\ at x = 0.15, and further to 15.59(3)~\AA\ in NGT (x = 1.0). The reduction of both the in-plane and out-of-plane lattice parameters upon Ni doping is associated with the transition of the pristine FGT crystal structure to another one that is similar to that of NGT, and has a smaller unit cell~\cite{deiseroth2006fe3gete2}. DFT calculations suggest that the contraction along the $c$-axis primarily originates from the presence of Ni atoms intercalated in the van der Waals (vdW) gap. Simulations of Ni-doped systems without vdW intercalation show minimal changes in the $c$-axis lattice parameter. In contrast, Ni intercalation between layers leads to step-like shrinkage of the $c$-axis, indicating the formation of covalent bonds between intercalated Ni atoms and neighboring Te atoms.
\begin{figure} [b!]
\centering
\includegraphics[width=0.9\linewidth]{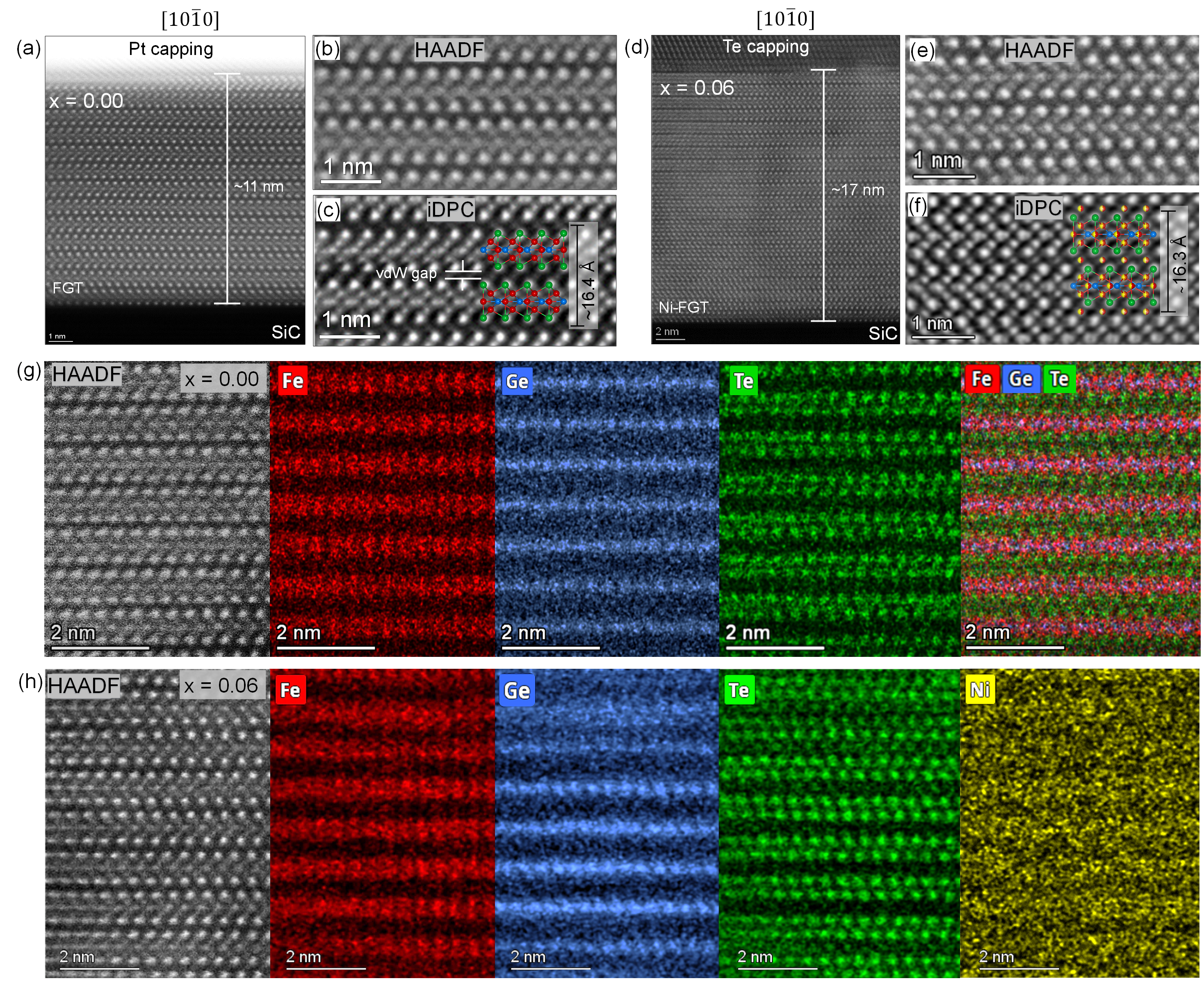}
\caption{\textbf{HAADF STEM and EDX mapping for pristine FGT and Ni-doped FGT film stacks} (a) Cross-sectional STEM image of 11 nm thick pristine FGT with a Pt capping, along $[10\Bar{1}0]$-direction. (b,c) HAADF and iDPC images confirm the formation of the FGT crystal structure with a d-spacing of 1.64 nm corresponding to the (0001) plane. (d) Cross-sectional STEM image for 17 nm thick Ni-doped (x = 0.06) FGT with a Te capping, along $[10\Bar{1}0]$-direction. (e,f) HAADF and iDPC image shows a d-spacing of 1.60 nm corresponding to the (0001) plane and also Ni intercalation into the vdW-gaps. (g,h) Cross-sectional EDX elemental mapping along $[10\Bar{1}0]$-direction for the pristine and Ni-doped FGT films, respectively. The mappings confirm the expected distributions for Fe, Ge, and Te, and reveal a homogeneous Ni distribution for Ni-doped FGT.}
\label{Fig:2} 
\end{figure}

\hspace{1cm}Scanning transmission electron microscopy (STEM) investigations using lamellae prepared by the focused ion beam (FIB) technique (see supplementary Fig. S5) confirmed the changes observed in the structure upon Ni doping. Fig.~\ref{Fig:2}a shows a cross-sectional high-angle annular dark-field (HAADF)-STEM image of 11 nm thick pristine FGT film with a Pt capping layer. The zoomed-in view in Fig.~\ref{Fig:2}b reveals the formation of an atomically flat and uniform layered structure, confirming the high-quality epitaxial growth of the FGT film on graphene/SiC(0001) with the vdW gaps well aligned parallel to the substrate surface. HAADF-STEM imaging relies on \textit{Z}-contrast mechanism, where intensity scales approximately with $Z^{1.7}$. This relatively high power in \textit{Z} makes the simultaneous visualization of light and heavy elements challenging. To overcome this limitation, we employed integrated Differential Phase Contrast (iDPC) imaging, which is about linear in \textit{Z} and provides excellent sensitivity when imaging light and heavy elements simultaneously. Fig.~\ref{Fig:2}c presents the iDPC image of the same region, clearly resolving the vdW gaps and all constituent elements. The image also shows QLs of FGT, similar to the atomic structure model in Fig.~\ref{Fig:1}a. Interestingly, in some regions of pristine FGT film, the signatures of irregular Fe intercalation were observed within vdW gaps between Te layers of FGT monolayers, as evidenced by atomic-resolution iDPC-STEM imaging (Supplementary Fig. S6), consistent with prior observations in MBE-grown Fe$_3$GeTe$_2$~\cite{wu2023fe} and Fe$_5$GeTe$_2$ films~\cite{silinskas_self-intercalation_2024}. We believe that this self-intercalation phenomenon is likely driven by stoichiometric excess of Fe in FGT during growth~\cite{wu2023fe}. Cross-sectional HAADF-STEM and iDPC-STEM images of a 17 nm thick Ni-doped (x = 0.06) FGT film with a Te capping layer are shown in Fig.~\ref{Fig:2}d,e and Fig.~\ref{Fig:2}f, respectively. In this case, the iDPC-STEM image clearly reveals that the addition of Ni leads to much more intercalation into the vdW gaps, leading to a reduction of the out-of-plane lattice parameter from 16.4~\AA~to 16.3~\AA, consistent with our XRD and DFT results. 

\hspace{1cm}In Fig.~\ref{Fig:2}g, an energy-dispersive X-ray spectroscopy (EDX) elemental map was obtained along $[10\overline{1}0]$ for pristine FGT film, which confirms the expected distribution of Fe, Ge, and Te, similar to the schematic Fig.
\ref{Fig:1}a. These EDX results agree well with the atomic arrangement in FGT layer observed in the HAADF-STEM images. For the Ni-doped (x = 0.06) FGT film, the EDX mappings were also obtained along the same $[10\overline{1}0]$ crystallographic direction [Fig.~\ref{Fig:2}h and Fig.~S5b]. It reveals a homogeneous spatial distribution of Ni, which suggests that its incorporation occurs both at Fe lattice sites in the unit cell and also within the vdW gaps. Therefore, TEM results confirm the intercalation effect during the growth, leading to the reduction of the lattice parameter. This is in accordance with our DFT calculations previously discussed.

\subsection*{Magnetic and anomalous Hall properties }\label{sec:3.3}
\begin{figure} [b!]
\centering
\includegraphics[width=0.9\linewidth]{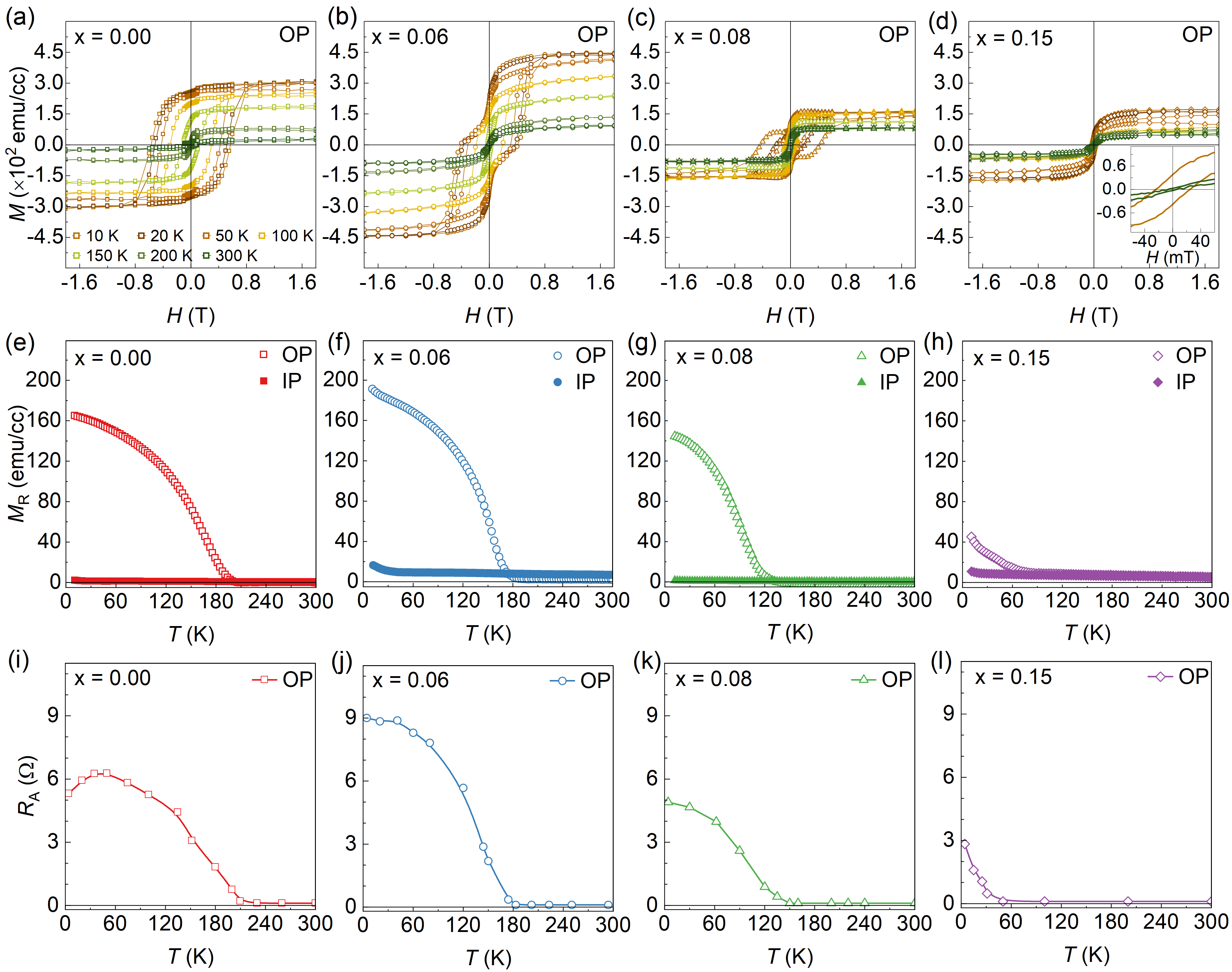}
\caption{\textbf{Magnetization and Hall measurements} (a-d) Magnetization (\textit{M}) versus applied magnetic field (\textit{H}) measured at $10-300$~K for (a) pristine FGT and (b-d) Ni-doped FGT films, by sweeping the \textit{H} in out-of-plane (OP) geometry. The inset shows a zoomed-in plot at 10 K and 300 K, revealing a finite $H_{\rm C}$ (25~mT) at 10 K. (e-h) Temperature-dependent magnetization remanence plot for all the films, after applying a large positive field $+5$~T to magnetize the film and set to 0~T before measuring in warming from $10-300$~K, in both in-plane (IP) and OP geometry, denoted by open and closed symbols, respectively. (i-l) Temperature-dependent anomalous Hall resistance $R_{\rm A}$ for all the films.}
\label{Fig:3}
\end{figure}
Figure~\ref{Fig:3}a–d depicts the magnetization hysteresis curves for all films measured at temperatures ranging from 10 to 300~K, in the out-of-plane (OP) geometry, where the magnetic field ($H$) is applied along the \textit{c}-axis. For pristine FGT (x = 0.00), we observe a well-defined hysteresis loop with a large coercive field ($H_{\rm C}$) of $\sim$0.6 T, similar to our previous reports~\cite{lopes2021large}. The saturation magnetization ($M_{\rm S}$) at 10 K is determined to be around 300 emu/cc ($\sim1.21~\mu_{\rm B}/{\rm Fe}$), comparable to that of bulk FGT crystals~\cite{drachuck2018effect} and MBE-grown FGT films~\cite{lopes2021large}. The Ni-doped FGT film (x = 0.06) presents a step-like hysteresis behavior which could be due to the co-existence of distinct magnetic domains~\cite{liu_wafer-scale_2017} or due to the intercalation effect~\cite{drachuck2018effect, zhu2024effect}. This feature persists in the (x = 0.08) Ni-doped film, though $M_{\rm S}$ is significantly reduced to 150 emu/cc ($\sim0.61~\mu_{\rm B}/{\rm Fe}$) at 10 K, indicating a reduction of PMA. These observations are further confirmed and analyzed in detail later by DFT calculations. For x = 0.15, the double loop is no longer present, and the coercivities further weaken, reaching a minimal $H_{\rm C}$ of 25~mT at 10~K as shown in the inset of Fig.~\ref{Fig:3}d, while the value of $M_{\rm S}$ is nearly unchanged, similar to previous results obtained for bulk crystals of Ni-doped FGT~\cite{drachuck2018effect} and Co-doped FGT.~\cite{chowdhury2021unconventional}
Hence, the results show that Ni doping $x\geq0.15$ results in a complete reduction of both magnetic exchange interaction and PMA, which we associate with Ni intercalation and Ni substitution at Fe sites. In contrast to Fe intercalation in FGT~\cite{wu2023fe} and Cr intercalation in CrTe$_2$~\cite{huang2021significant}, where an enhancement of overall $T_{\rm C}$ was observed, we found that the Ni intercalation reduces $T_{\rm C}$ due to the non-magnetic nature of the Ni atom in the FGT structure.  We attribute this decrease in $T_{\rm C}$ to excess Ni incorporation, both as intercalation in the vdW gaps and as substitutions at Fe sites, as previously shown by HAADF-EDX map in Fig.~\ref{Fig:2}f, similar to Ni-doped in bulk FGT~\cite{drachuck2018effect} and FGaT crystals~\cite{yuan2025direct, zhu2024effect}. 

\hspace{1cm}Compare to bulk crystals~\cite{drachuck2018effect, zhu2024effect}, thin films exhibit a larger $H_{\rm C}$, making them promising for stabilizing robust PMA with a finite remanent magnetization ($M_{\rm R}$) that persists up to $T_{\rm C}$. Fig.~\ref{Fig:3}e–h shows the temperature-dependent $M_{\rm R}$ measured from 10 K to 300 K for all films in both OP and IP configurations. For the pristine FGT film, a dominant $M_{\rm R}$ contribution along the OP direction, with negligible IP remanence, confirms robust PMA. $T_{\rm C}$ is found to be around 210~K, similar to previous reports~\cite{lopes2021large, liu_wafer-scale_2017, tan2018hard, may2016magnetic}. For Ni-doped FGT [see Fig.~\ref{Fig:3}f-h], we observed a drastic reduction of $T_{\rm C}$ from 210~K (for pristine FGT) down to 50~K (Ni-doped FGT, x = 0.15). The reduction of $T_{\rm C}$  upon Ni incorporation suggests a weakening of long-range ferromagnetic ordering due to structural deformation of the unit cell. To conclude, from the SQUID measurement, we observed that $M_{\rm S}$, $M_{\rm R}$, and $H_{\rm C}$ decrease progressively with increasing temperature. 

\hspace{1cm}To further validate the transition of $T_{\rm C}$ on Ni incorporation, temperature-dependent anomalous Hall effect (AHE) measurements were systematically performed on all films (see Supplementary Fig~S7). AHE hysteresis loops were acquired by sweeping an external magnetic field ($\pm0.8$~T) applied perpendicular to the film surface~\cite{nagaosa2010anomalous, shinwari2025above, khan2022intrinsic}. Fig.~\ref{Fig:3}i–l displays the extracted anomalous Hall resistance ($R_{\rm A}$) as a function of temperature for pristine FGT and Ni-doped FGT films. For the pristine FGT film, $T_{\rm C}$ is determined to be approximately 216~K, which aligns well with the value extracted from our SQUID measurements. Notably, the value of $T_{\rm C}$ exhibits a systematic reduction with increasing Ni doping concentration, further corroborating the SQUID-derived results shown in Fig.~\ref{Fig:3}e–h. The reduction of $T_{\rm C}$ trend confirms the reduction of ferromagnetic ordering upon Ni incorporation, suggesting a suppression of the magnetic exchange interactions in the Ni-doped FGT system, in agreement with the DFT calculations discussed later. We also measured the AHE signal for FGT films with different capping layers (Pt and Te), confirming the negligible role of capping layer in modulating the $T_{\rm C}$ of the ferromagnetic FGT film (see Fig.~S8a,b).

\subsection*{X-ray magnetic circular dichroism (XMCD)}
\begin{figure} [t!]
\centering
\includegraphics[width=0.9\linewidth]{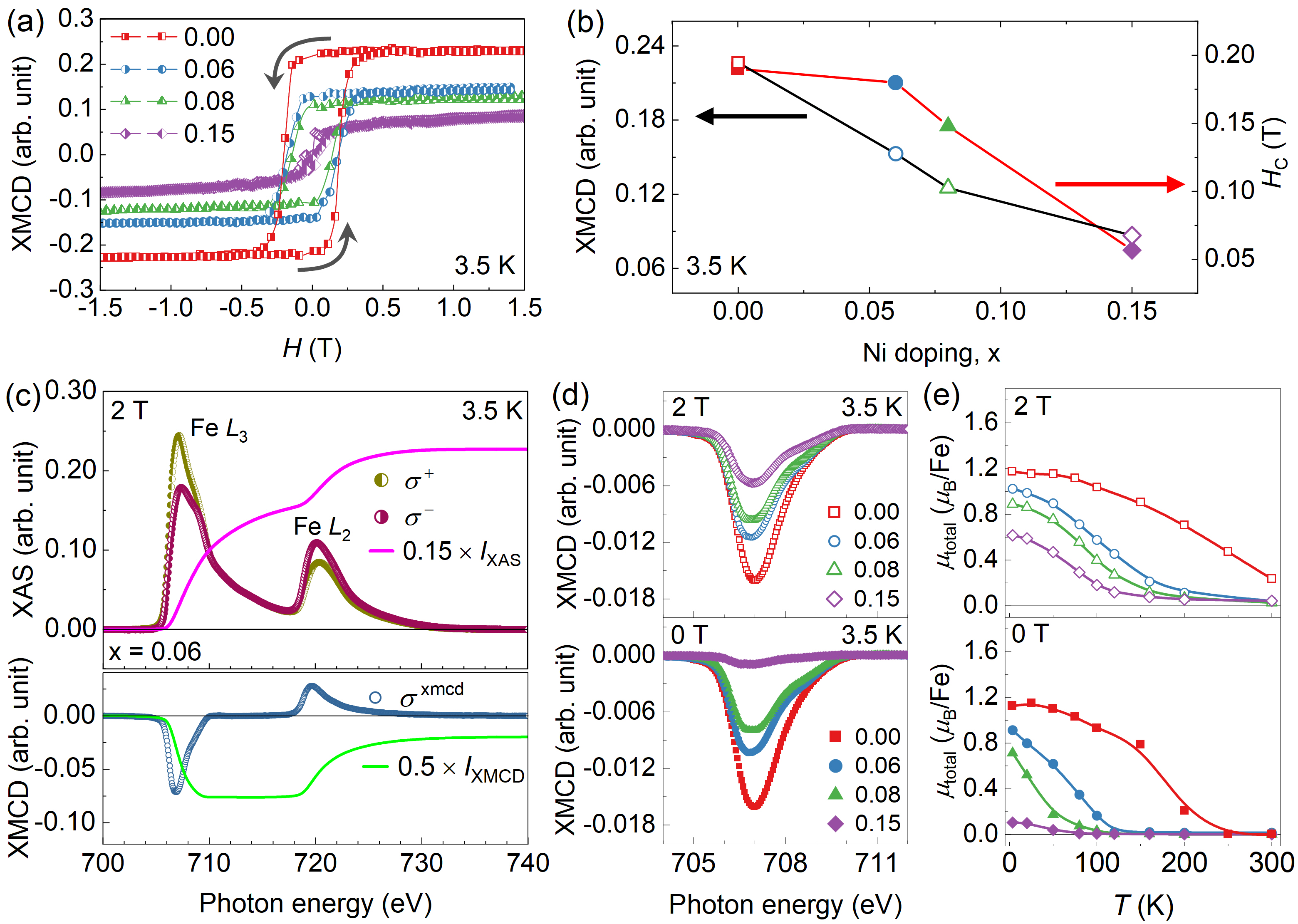}
\caption{\textbf{XAS and XMCD for pristine FGT and Ni-doped FGT film stacks} 
(a) XMCD hysteresis loops measured at 3.5 K for pristine and Ni-doped FGT films. (b) Dependence of XMCD signal amplitude and coercive field ($H_{\rm C}$) with Ni doping concentration. (c) XAS (open symbols) and XMCD (closed symbols) spectra for Ni-doped (x = 0.06) FGT film, with solid lines showing the integrated signals. (d) Comparison of XMCD spectra (Fe $L_3$-edge) acquired at 0~T and 2~T for all compositions. (e) Temperature dependence of the total magnetic moment ($\mu_{\rm total}$) in remanence (0~T, closed symbols) and saturation field (2~T, open symbols).} 
\label{Fig:4} 
\end{figure}
XMCD measurements were performed at the Fe $L_3$ and $L_2$-edges of FGT films under ultra-high-vacuum (UHV) sample environment (~$1\times10^{-10}$ mbar) with a temperature ranging from 3.5~K to 300~K. Unlike SQUID magnetometry, which measures the net macroscopic magnetization of the film, XMCD is surface sensitive with soft X-ray and total electron yield (TEY) dectection. This technique is extremely useful to provide element-specific magnetic information, as well as the possibility to disentangle spin and orbital moments, which is crucial for understanding doping-induced changes in ferromagnetic materials. Fig.~\ref{Fig:4}a, shows the XMCD hysteresis loop at 3.5 K for pristine and Ni-doped FGT films, acquired at the Fe $L_3$-edge (706.9 eV) and normalized by the pre-edge region at 700 eV. In Fig.~\ref{Fig:4}b, the observed monotonic decrease in the XMCD signal and its corresponding $H_{\rm C}$ with Ni doping clearly demonstrates the weakening of PMA, corroborating the trends observed by SQUID and Hall. The X-ray propagation direction was aligned parallel to both the sample normal and the applied magnetic field. The XAS spectra [$(\sigma^+ + \sigma^-)/2$] for all samples at the Fe and Ni $L_{2,3}$-edges, measured at 3.5~K, are shown in Fig~S9a,b. The presence of the Ni edge in the doped films further validates the Ni incorporation. For all Ni-doped films, although the Ni $L_{2,3}$-edge XAS signal was measurable, it was not strong enough to reveal any clear trend in the XMCD spectra at these edges, likely due to both low atomic concentration of Ni atoms and its associated low magnetic moment.
\begin{table}[htbp!]
\caption{The extracted values of orbital and spin magnetic moments measured at 3.5~K using Fe $L_3$-edges for pristine and Ni-doped FGT films.}
\centering 
\setlength{\tabcolsep}{20pt}
\renewcommand*{\arraystretch}{1}
\begin{tabular}{c|c|c|c|c}
\hline
\hline
Doping & $\mu_{\rm spin}$ & $\mu_{\rm orb}$ & $\mu_{\rm total}$ & $\mu_{\rm orb}/\mu_{\rm spin}$\\
 x  &  $\mu_{\rm B}/{\rm Fe}$ & $\mu_{\rm B}/{\rm Fe}$ &  $\mu_{\rm B}/{\rm Fe}$ & \\
\hline 
0.00   & 1.130 & 0.047 & 1.180 & 0.042 \\ 
0.06   & 1.016 & 0.067& 1.074 &  0.066\\  
0.08   & 0.903 & 0.062 & 0.966 & 0.069 \\ 
0.15   & 0.625 & 0.023 & 0.649 &  0.037\\ 
\hline
\hline
\end{tabular}
\label{tab1}
\end{table}

\hspace{1cm}Fig.~\ref{Fig:4}c presents the Fe $L_{2,3}$-edges XAS (upper panel) and XMCD (lower panel) spectra only for the Ni-doped (x = 0.06) FGT film, measured at an applied field of 2 T. The integrated intensities of the XAS ($I_{\rm XAS}$) and XMCD ($I_{\rm XMCD}$) signals are also calculated and displayed in the figure. Similar spectral features were obtained for all samples. Fig.~\ref{Fig:4}d compares the XMCD spectra at Fe $L_3$-edge with 0~T and 2~T for all films. By applying the XMCD sum rule analysis~\cite{thole1992x,carra1993x}, we quantitatively determined the orbital ($\mu_{\rm orb}$), spin ($\mu_{\rm spin}$), and total ($\mu_{\rm total}$) magnetic moments at 3.5 K for all films (see Table~\ref{tab1}), with estimated uncertainties of $\Delta(\mu_{\rm orb}) = 0.006~\mu_{\rm B}/{\rm Fe}$, $\Delta(\mu_{\rm spin}) = 0.120~\mu_{\rm B}/{\rm Fe}$, and $\Delta(\mu_{\rm total}) = 0.130~\mu_{\rm B}/{\rm Fe}$.  The temperature dependence of the total magnetic moment ($\mu_{\rm total}$) in both remanence (0 T) and saturation (2 T) states is shown in Fig.~\ref{Fig:4}e. For pristine FGT (x = 0.00), we found $\mu_{\rm total} = 1.180~\mu_{\rm B}/{\rm Fe}$ at 3.5~K, consistent with SQUID data. For Ni doping (x = 0.15), $\mu_{\rm total}$ reduces to 0.649~$\mu_{\rm B}/{\rm Fe}$, primarily through suppression of $\mu_{\rm spin}$, reflecting the weakening of long-range ferromagnetic order upon Ni incorporation.

\begin{figure} [b!]
\centering
\includegraphics[width=0.9\linewidth]{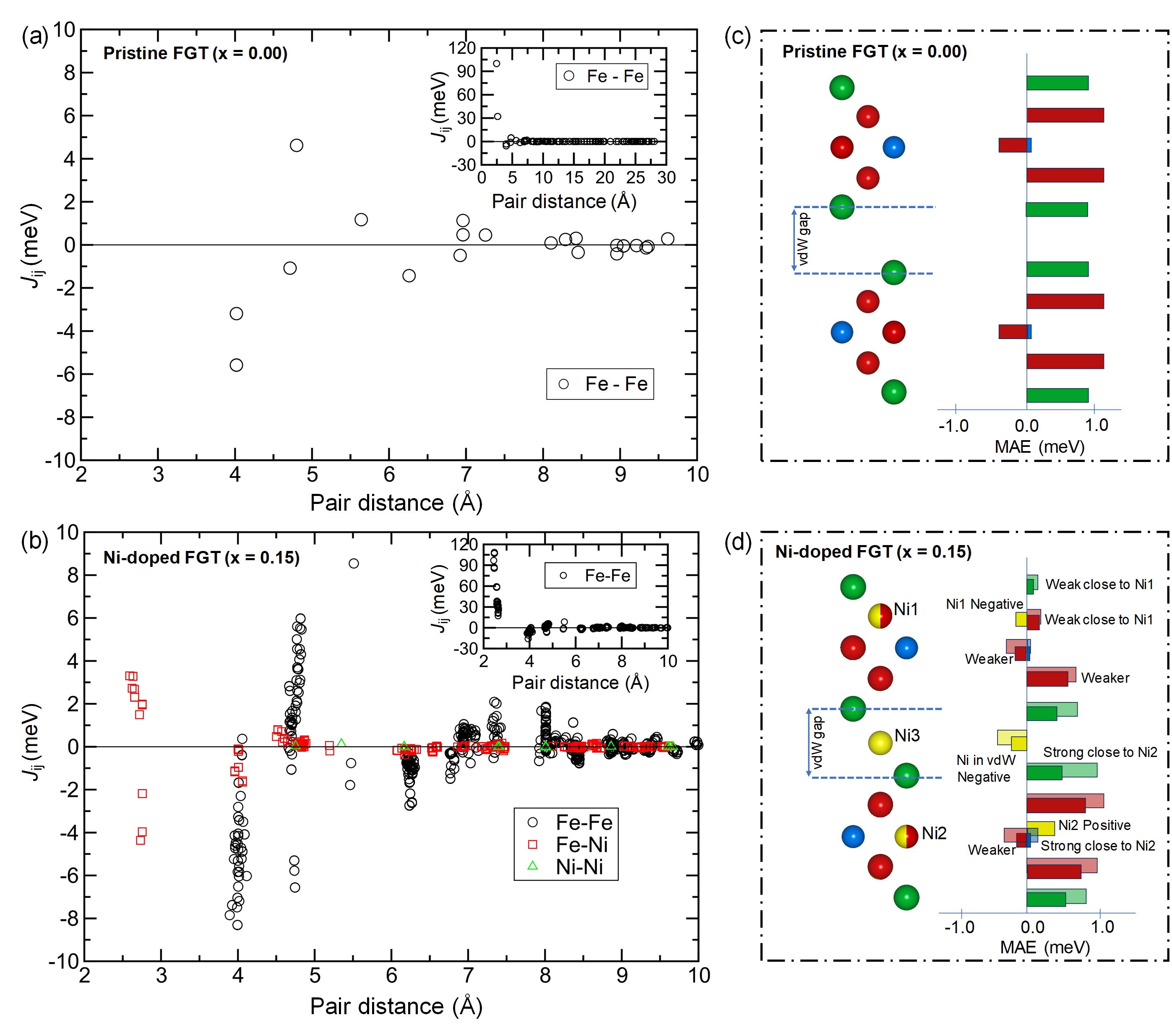}
\caption{\textbf{DFT results on magnetism.} Calculated exchange interaction parameters $J_{\rm ij}$ (truncated between -10 to +10 meV) for (a) Fe-Fe pairs in pristine FGT (x = 0.00), (b) Fe-Fe, Fe-Ni and Ni-Ni pairs in Ni-doped FGT (x = 0.15). Multiple data points for every neighbor interaction shown in (b) correspond to various pair-exchange parameters in the supercell. In both (a) and (b), the insets show all the values of Fe-Fe exchange interaction parameters. Atomic structure and its atom-projected magnetocrystalline anisotropy energies (MAE) are shown for (c) pristine FGT, (d) Ni-doped FGT. The colors of the bars shown on the right side correspond to the colors of the atoms shown in the structures on the left side. In (d), the mixed colored balls indicate the sublattice sites shared between Fe and Ni atoms for which the corresponding atom-projected anisotropies are shown.} 
\label{Fig:5} 
\end{figure}
\section*{Density Functional Theory (DFT)}
Based on our DFT calculations, the interatomic exchange parameters ($J_{\rm ij}$) presented in Figs.~\ref{Fig:5}a,b, indicate a complex behavior with Ni doping case compared to pristine FGT. It is clearly observed that Ni doping introduces strong antiferromagnetic (AFM) coupling between Fe-Fe and Fe-Ni pairs, which compete with dominant ferromagnetic (FM) Fe-Fe interactions. Moreover, Ni-Ni exchange parameters are found out to be very weak. Therefore, the substitution of Fe by Ni overall decreases the FM component. It should also be noted that Ni doping introduces structural distortions, which are reflected in slightly different values of the pair distance for similar neighbor interactions shown in Fig.~\ref{Fig:5}b. The corresponding exchange interaction parameters also vary in magnitude and sometimes in sign. Figs.~\ref{Fig:5}c,d show calculated atom-projected magnetocrystalline anisotropy energies (MAE) for pristine and 15 \% doped Ni systems respectively. In these figures, positive (+ve) values denote contributions of MAE favoring out-of-plane (perpendicular) magnetic anisotropy and negative (-ve) values denote contributions favoring in-plane magnetic anisotropy. In both cases, the easy axis of magnetization is perpendicular to the plane as observed in experiments. However, Ni doping significantly decreases PMA (0.61 meV/atom for pristine in contrast to 0.34 meV/atom for Ni-doped case). For the pristine system, almost all the atoms contribute towards PMA while for the doped system, several factors decrease the overall PMA. Firstly, Ni in vdW gap shows an in-plane magnetization as evident from the -ve value of the atom-projected magnetocrystalline anisotropy energy. However, its effect on the neighboring Fe atoms is less significant, whereas Ni substituted at Fe sites strongly suppresses the PMA of neighboring Fe atoms. All these effects act together to decrease the overall PMA of the FGT structure by a significant amount. Therefore, consistent results from SQUID, Hall, XMCD, and DFT demonstrate that Ni-incorporation in FGT films suppresses the magnetic properties, including PMA and reduces the $T_{\rm C}$ down to 50~K (see supplementary Fig.~S10), due to modification in FGT crystal structure. 

\section*{Conclusion}
In conclusion, we demonstrate for the first time the effects of Ni doping on the structural and magnetic properties of large-area FGT films epitaxially grown on graphene templates via MBE. Structural characterization via different methods confirms their high crystalline quality. A reduction of both in-plane and out-of-plane lattice parameters was observed for the Ni-doped films, suggesting unit cell contraction due to intercalation effects, an effect also confirmed by DFT calculations. Magnetization, Hall, and XMCD measurements confirm the ferromagnetic nature with robust PMA for pure FGT as well as for all Ni-doped films. However, for Ni-doped FGT films, $T_{\rm C}$ decreases with Ni incorporation. We found that Ni doping modifies the lattice parameters (e.g., Ni intercalation/substitution), which results in the reduction of magnetic properties of FGT due to the suppression of ferromagnetic ordering and decrease in $T_{\rm C}$ from 210~K (x = 0) down to $\sim$ 50 K (x = 0.15). Our DFT calculations of interatomic magnetic exchange parameters and atom-projected magneto-crystalline anisotropy energies have provided fundamental insights to understand the effects of Ni doping. Furthermore, both XMCD and DFT results reveal that the reduction in $\mu_{\rm spin}$ plays a major role in suppressing the ferromagnetic properties. The present study not only shows how intercalation effects play a crucial role in the magnetic properties, but also demonstrates the feasibility of realizing controlled synthesis of high-quality 2D ferromagnetic materials. This is crucial for the development of devices exhibiting tailored magnetic and electronic properties for future spintronic applications based on 2D magnets and heterostructures.

\section*{Methods}
\subsection*{Thin film growth}
We prepared high-quality FGT and Ni-doped FGT films using MBE. Similar to our previous works~\cite{lopes2021large, shinwari2025above}, we chose as substrates epitaxial graphene grown on semi-insulating 4H-SiC(0001) using the SiC surface graphitization method ~\cite{heilmann2020influence}. Large-area FGT and Ni-doped FGT films were prepared with a growth rate of 0.55~nm/min at a substrate temperature of 300~$^\circ{\rm C}$. In order to obtain the desired stoichiometry of Fe$_3$GeTe$_2$ phase, an optimized flux ratio rate (or beam equivalent pressure - B.E.P. ratio) of 1:1:21-24 (Fe:Ge:Te) was employed. In order to control the Ni-doping in FGT films, the flux of Fe and Ni effusion cells were calibrated by monitoring the B.E.P. ratios and RBS measurements. The base pressure of the growth chamber was maintained below 1$\times$10$^{-10}$ mbar, while the pressure during growth was about 1$\times$10$^{-9}$ mbar. To monitor and confirm the epitaxial growth of the films, \textit{in-situ} RHEED was used during the whole growth process. After the growth of the desired thickness of film, a thin capping layer (5-7~nm) of either Pt or Te was deposited \textit{in-situ} at room temperature, to protect the films from oxidation upon air exposure. The Te capping was deposited using an effusion cell, while Pt was deposited via \textit{in-situ} magnetron sputtering.
\\
\\
\textbf{Focused Ion Beam (FIB) sample preparation:} The FEI Helios G4 CX dual-beam SEM FIB (scanning electron microscope focused ion beam) was used to prepare electron-transparent samples. The protective Electron Beam Induced Deposition (EBID) carbon (C) and Pt were deposited on the film before the Ion Beam Induced Deposition (IBID) Pt protective layer (see supplementary Fig.~S5a). Cross-sectional chunks (dimensions: $15\times2.0\times5~{\rm \mu m}^3$) were made and transferred using an EasyLiftTM needle to the copper half-grid. Then the sample was initially thinned down to 80-100 nm thickness using standard Ga-beam processing at 30 kV within a certain window ($6.0\times5.0~{\rm \mu m}^2$). The remaining chunk is left thick to have a rigid frame to minimize the bending and stress release in the e-transparent window. Finally, several low kV cleaning steps (5 and 2 kV) were used to clean the side surfaces of the lamellae. 
\\

\subsection*{Structural characterization}
\textbf{Rutherford Backscattering Spectrometry (RBS):} The chemical composition of Fe, Ni, Ge, and Te in all films was obtained by the Rutherford backscattering, with the help of 1.7~MeV He$^{+}$ ions at a scattering angle of 170$^\circ$. By fitting the RBS spectra with the program ndf v9.3g~\cite{barradas1997simulated}, the doping ratio of Ni/Fe were determined to be 0.00, 0.06, 0.08, and 0.15 (see supplementary file S1).

\textbf{X-Ray Diffraction (XRD):} To investigate the structural phase formation and determine the out-of-plane lattice parameters, $\theta-2\theta$ scans were performed using a PANalytical X'Pert Pro MRD diffractometer with Cu-$K_{\alpha1}$ radiation ($\lambda$ = 1.54056 \AA). To probe the in-plane crystallographic properties of the layers, such as the in-plane lattice parameters and azimuthal orientation, grazing-incidence diffraction (GID) was performed. GID experiments were conducted at the BM25B SpLine beamline at the European Synchrotron Radiation Facility (ESRF) using X-rays with an photon energy of 18 keV. GID is an extremely surface-sensitive analytical technique for the aforementioned quantities because the X-ray wave field exponentially diminishes in depth when the angle of incidence is close to the critical angle of total external reflection ($\alpha_{\rm C}= 0.12^\circ$). Here, we chose a slightly larger angle of 0.15° as a compromise, in order to obtain a sufficient signal from the substrate (which is necessary for referencing the layer orientation), while remaining highly surface sensitive. 
\\
\\
\textbf{Scanning Transmission Electron Microscopy (STEM):} The FIB-prepared and loaded TEM grid was transferred immediately to the TEM column using a dedicated double-tilt TEM holder optimized to collect X-rays in the TEM. The microstructure of samples were examined with a double-corrected (probe and image correctors) and monochromated Themis Z scanning transmission electron microscope (Thermo Fisher Scientific) operating at 300 kV. The STEM images were acquired through HAADF (high-angle annular dark-field) mode and simultaneously integrated differential phase contrast (iDPC) mode. The beam convergence angle was measured at $\sim$24.5 mrad, and the probe current of 15-25 pA was used for STEM imaging. Energy dispersive X-ray spectroscopy (STEM EDS maps) results were achieved with a Dual X EDS system (Bruker) using two large area detectors in total, capturing 1.76 steradian with a probe current of 25 pA for more than 2 hours. Data acquisition and analysis were done using Velox software.

\subsection*{Magnetic and Hall measurements}
\textbf{Superconducting Quantum Interference Device (SQUID):} 
The static magnetic hysteresis loops were obtained in both in-plane and out-of-plane geometry, using a SQUID-based MPMS3 5XL by Quantum Design Inc. To determine the temperature-dependent remanence magnetization ($M_{\rm R}$) for all films, we first saturated the magnetic moments along the out-of-plane (OP) direction by applying a strong magnetic field ($H=5$~T) at 10~K  and then reduced it to zero ($H=0$~T) in linear mode. The same procedure was repeated for the in-plane (IP) configuration for all the samples to extract the $M_{\rm R}$.
\\
\\
\textbf{Hall setup:} 
The Hall measurements were performed in high-vacuum environment of 10$^{-6}$~mbar, on a Hall setup,  using a nanovoltmeter and a current source meter. An electromagnet was used to sweep the applied magnetic field from +0.8~T to -0.8~T. We used a sample of dimension 5~mm $\times$ 10~mm on a chip carrier and bonding it with Al wires as contact terminals. One pair of contact terminals was used for applied DC current ($I_{xx}$) and another pair to measure the measured voltage drop ($V_{xy}$) towards the orthogonal direction to the applied current in the film plane. 

\subsection*{X-ray Absorption Spectroscopy (XAS) and X-ray Magnetic Circular Dichroism (XMCD).}
XAS/XMCD measurements were performed at the Fe $L_3$ and $L_2$ edges using the BOREAS (BL29) beamline~\cite{barla2016design} located in ALBA synchrotron (Spain). The XAS signal was obtained by total-electron-yield detection, where the drain current was measured from the sample to the ground. Furthermore, the XMCD spectrum ($\sigma^{\rm xmcd}$) was obtained from the difference between the left ($\sigma^-$) and right ($\sigma^+$) polarized X-ray spectra. The total magnetic moment is obtained by summing, $\mu_{\rm total} = \mu_{\rm orb} + \mu_{\rm spin}$ where the orbital ($\mu_{\rm orb}$) and spin ($\mu_{\rm spin}$) magnetic moments were obtained by calculating the XMCD magneto-optical sum rules~\cite{thole1992x, carra1993x} applied to the Fe $L_{2,3}$ edges~\cite{chen_experimental_1995}. The magnetic hysteresis loop was recorded for both pristine FGT and Ni-doped FGT films at 3.5 K with the X-ray photon energy fixed at the maximum of the Fe $L_3$ XMCD signal at 706.9 eV normalized by the pre-edge region at 700~eV~\cite{goering2000element}, measured on the fly with a magnetic field ramp of 2 T/min at high applied magnetic field (between -2 T and -1 T, between 1 T and 2 T) and of 0.2 T/min at low applied magnetic field (between -1 T and 1 T) for two circular polarizations.

\subsection*{Density Functional Theory (DFT) calculations}
First-principles structural optimizations, formation-energy evaluations, and magnetic-property calculations were conducted within the density functional theory (DFT) framework  \cite{kohn1965self,hohenberg1964inhomogeneous} using the Vienna \textit{Ab initio} Simulation Package (VASP) \cite{kresse1993ab,kresse1994ab,kresse1996efficiency,kresse1996efficient}. The exchange-correlation potential was treated using the generalized gradient approximation (GGA) functional in conjunction with the Perdew, Burke, and Ernzerhof (PBE) method \cite{perdew1996generalized}. The projector augmented wave method \cite{Blochl1994projector} was applied. A plane-wave basis set with a kinetic cutoff energy of 500 eV was used to expand the electronic wave function. During structural optimization, the maximum force on each atom was less than 1$\times$10$^{-2}$ eV/\AA~. A~Gaussian smearing factor of 0.1 eV was taken into account. Brillouin zone (BZ) integration was performed by a $\Gamma$-centered $11\times11\times3$ uniform $k$-point grid for the unit cell and $5\times5\times3$ for the $2\times2\times1$ supercell. To account for vdW interactions, the DFT-D3 correction of Grimme with zero damping (IVDW = 11)~\cite{grimme2010consistent} was included in all calculations. Formation energies ($E_{f}$) and various doping configurations were evaluated in the $2\times2\times1$ supercell. The substitutional formation energies was calculated relative to pristine FGT, body-centered cubic (BCC) Fe, and face-centered cubic (FCC) Ni according to the following relations:
\[
E_{f(\text{Ni})}
= E_{t(\text{doped})}
- \left[ E_{t(\text{pristine})} - E_{\text{bcc-Fe}} + E_{\text{fcc-Ni}} \right].
\]
Magnetic properties (magnetic moments, Heisenberg exchange interaction, Dzyaloshinskii–Moriya interaction, and magnetic anisotropy energy) were calculated via the QuantumATK-Synopsys package version U-2022, employing an LCAO basis set, the ``PseudoDojo" pseudopotential \cite{qatk1, qatk2}, and a density mesh cutoff of 120 Hartree.

\section*{Acknowledgement}
The authors would like to thank H.-P. Schönherr, C. Herrmann, and C. Stemmler for their dedicated maintenance of the MBE system. The authors also appreciate the critical reading of the manuscript by Dr. Philipp John. XAS/XMCD measurements were carried out at the BOREAS beamline at ALBA synchrotron, Spain (proposal number 2024028287). The RBS measurements were carried out at IBC  at the Helmholtz-Zentrum Dresden-Rossendorf e.V., a member of the Helmholtz Association. They also acknowledge the provision of beamtime under the project HC-5796 at the European Synchrotron Radiation Facility (ESRF), located in Grenoble (France) and the PHARAO station at BESSY II, Helmholtz-Zentrum Berlin. B.S. acknowledges financial support from Swedish Research Council (grant no. 2022-04309) and STINT Mobility Grant for Internationalization (grant no. MG2022-9386). The computational resources provided by the National Academic Infrastructure for Supercomputing in Sweden (NAISS) at NSC and PDC (NAISS 2025/3-68) partially funded by the Swedish Research Council through grant agreement no. 2022-06725 and at UPPMAX (UPPMAX 2025/2-468) are gratefully acknowledged. 
\subsection*{Conflict of interest}
The authors declare no competing interests.

\section*{Authors' contributions} 
J.M.J.L. proposed and coordinated the project. K.I.A.K. executed the project and grew the samples. K.I.A.K., T.S., and M.H. performed structural characterization. A.M. and B.K. performed and analysed the high-resolution transmission electron microscopy data. K.I.A.K., T.S., H.L., and J.H. contributed to the magnetic and Hall measurements. F.M. measured and simulated the Rutherford backscattering spectroscopy data. W.L., K.I.A.K., A.I.F., M.V., S.R.G., and L.A. contributed to the X-ray magnetic circular dichroism measurement. S.E. and B.S. performed the density functional theory calculations. K.I.A.K. prepared the manuscript draft. All authors contributed to data analysis, discussion, and co-wrote and revised the manuscript.

\setcounter{figure}{0}  
\setcounter{table}{0}  
 \renewcommand{\thefigure}{S\arabic{figure}}
\renewcommand{\theequation}{S\arabic{equation}}
\renewcommand{\thetable}{S\arabic{table}}
\renewcommand{\thepage}{S\arabic{page}}

\renewcommand{\thesection}{S\arabic{section} :}
\setcounter{section}{0}
\renewcommand{\thesubsection}{S\arabic{subsection}}
\setcounter{subsection}{0}
\section*{\Large \centering Supporting Information for: Tailoring Magnetism and Structure in Large-Area Epitaxial Fe$_3$GeTe$_2$ via Ni-Doping}
\section{Rutherford backscattering spectrometry (RBS)}
\begin{figure*} [htbp]
\centering
\includegraphics*[width=0.9\linewidth]{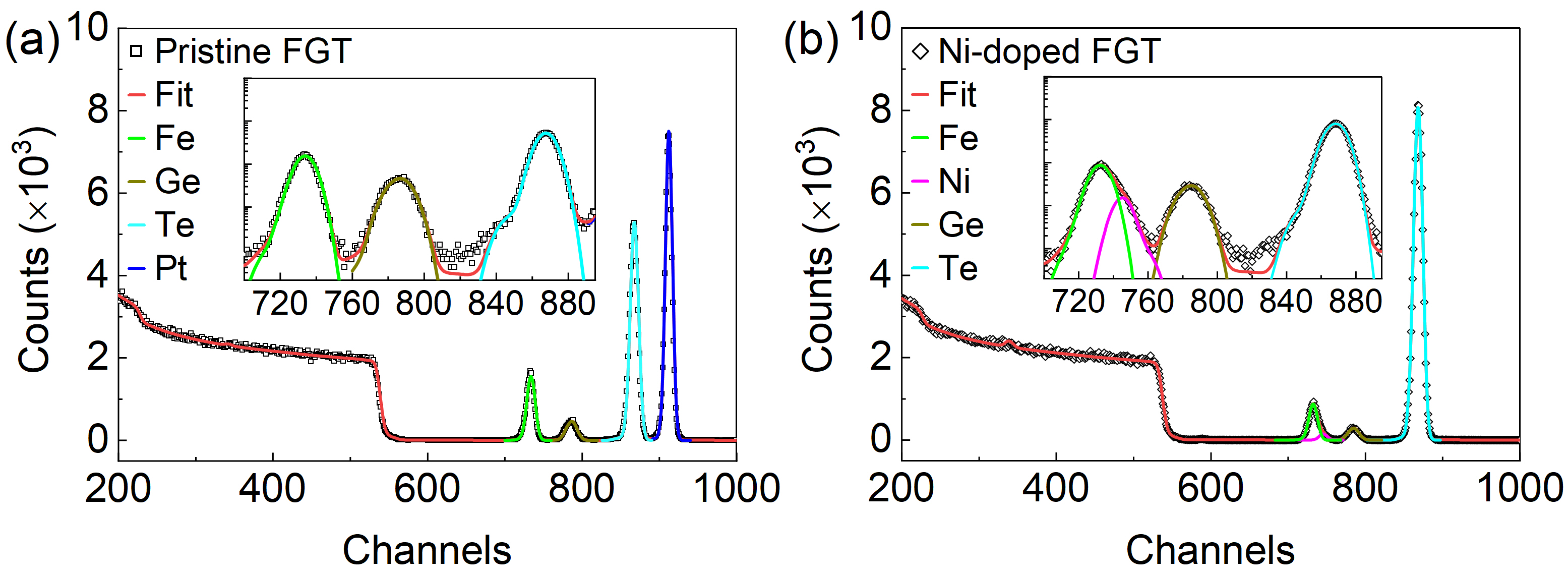}
\caption{\label{fig:RBS} (a,b) RBS spectra for pristine Fe$_3$GeTe$_2$ and Ni-doped Fe$_3$GeTe$_2$ films. The insets represent the zoomed spectra corresponds to Fe, Ni, Ge, and Te. The actual data and simulated fits are denoted by symbols and solid lines, respectively. The peak fitted with blue color for pristine FGT film corresponds to the Pt capping layer.}
\end{figure*}

Rutherford backscattering spectrometry (RBS) spectra were used to calibrate the chemical composition of Fe, Ni, Ge, and Te in pristine Fe$_3$GeTe$_2$ and Ni-doped [Fe$_{1-x}$Ni$_x$]$_3$GeTe$_2$ films. Figure~\ref{fig:RBS}(a,b) shows the RBS spectra for [Fe$_{1-x}$Ni$_x$]$_3$GeTe$_2$ films, where the ratio of Ni/Fe `x' was controlled during deposition and optimized by monitoring the beam equivalent pressure (B.E.P.) ratios of the Ni and Fe effusion cell fluxes. 
\newpage
\section{Preferential occupation for Ni-doping.}
\begin{figure*} [htbp!]
\centering
\includegraphics*[width=0.9\linewidth]{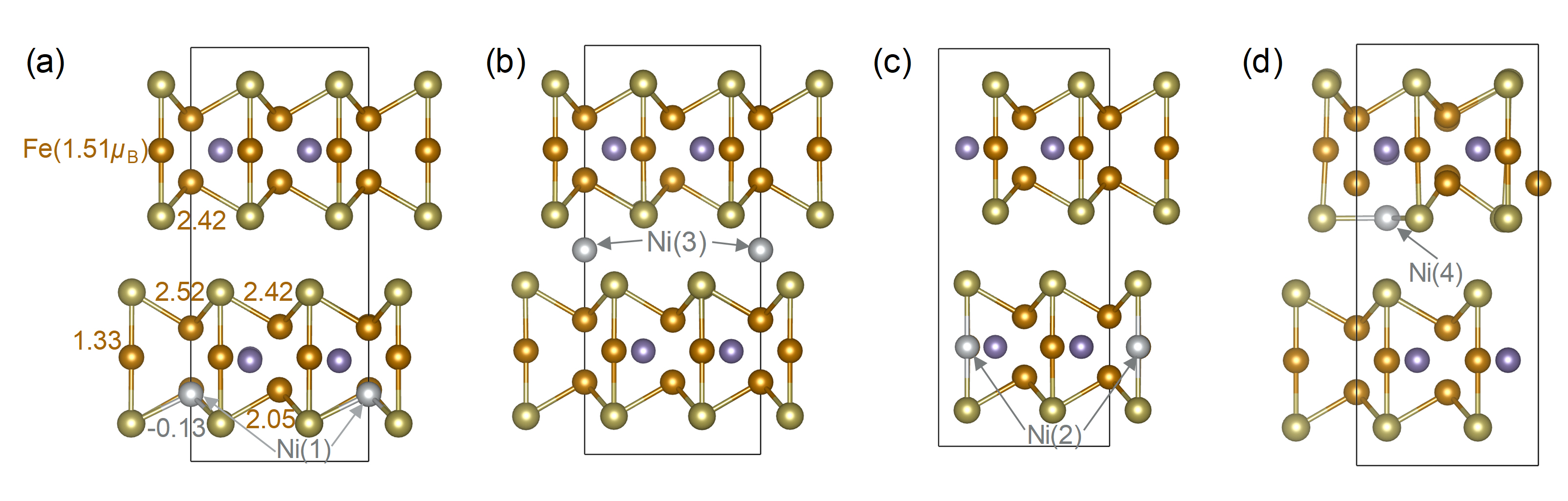}
\caption{\label{fig:Ef}  FGT structure with energetically favorable  configurations by incorporating Ni atoms at different Fe sides at a) outer Ni(1) sublattice, b) Ni(3) vdW-gaps, c) inner Ni(3) sublattice and, Ni(4) aligned with Te/Ge.}
\end{figure*}
As illustrated in Figure~\ref{fig:Ef}a-d, Ni doping in FGT can occur in multiple configurations. Ni atoms may partially substitute Fe atoms in the Fe(1) and Fe(2) sublattices, denoted as Ni(1) and Ni(2), respectively, or occupy interstitial positions within the vdW gap aligned with the Fe(1) layer, referred to as Ni(3). DFT calculations of formation energies at a Ni concentration of $4.17\%$ (see Table~\ref{tab_Ef}) reveal that substitution at the Ni(1) site is energetically favorable, with a formation energy of $E_{\rm f}$ = -0.11 eV/atom. The interstitial Ni(3) incorporation within the vdW gap is also energetically favorable ($E_{\rm f}$= -0.02 eV/atom). The substitution at the Ni(2) site requires an additional energy of 0.23 eV/atom. The Ni occupation in an alternative interstitial site above the Ge layer (Ni(4)) is not stable, with $E_{\rm f}$= 0.34 eV/atom. In contrast to pristine NGT, Ni atoms in Ni-doped FGT possess small spin magnetic moments ranging from 0.01 to -0.25 $\mu_{\rm B}$, depending on their specific occupation sites. For Ni atoms occupying the Ni(1) (-0.13 $\mu_{\rm B}$) and Ni(3) (-0.25 $\mu_{\rm B}$) positions, the magnetic moments are antiparallel to those of the surrounding Fe atoms, resulting in a weak ferrimagnetism within the structure.

\begin{table}[htbp!]
\caption{Ni-doping ($4.17\%$) in FGT through multiple configurations depending on formation energies and its spin magnetic moment contribution.}
\centering 
\setlength{\tabcolsep}{25pt}
\renewcommand*{\arraystretch}{1.4}
\begin{tabular}{l|c|c}
\hline
\hline
Lattice sites & Formation energies & Spin magnetic moment  \\
 & $E_{\rm f}$ (eV/atom) & ($\mu_{\rm B}$) \\
\hline 
Ni(1)   & -0.11  & -0.13\\  
Ni(3)   & -0.02 & -0.25 \\ 
Ni(2)   & 0.23 & 0.11  \\ 
Ni(4)   & 0.34 & 0.01  \\ 
\hline
\end{tabular}
\label{tab_Ef}
\end{table}
\newpage
\section{Grazing-incidence diffraction (GID) }
\begin{figure*} [b!]
\centering
\includegraphics*[width=0.8\linewidth]{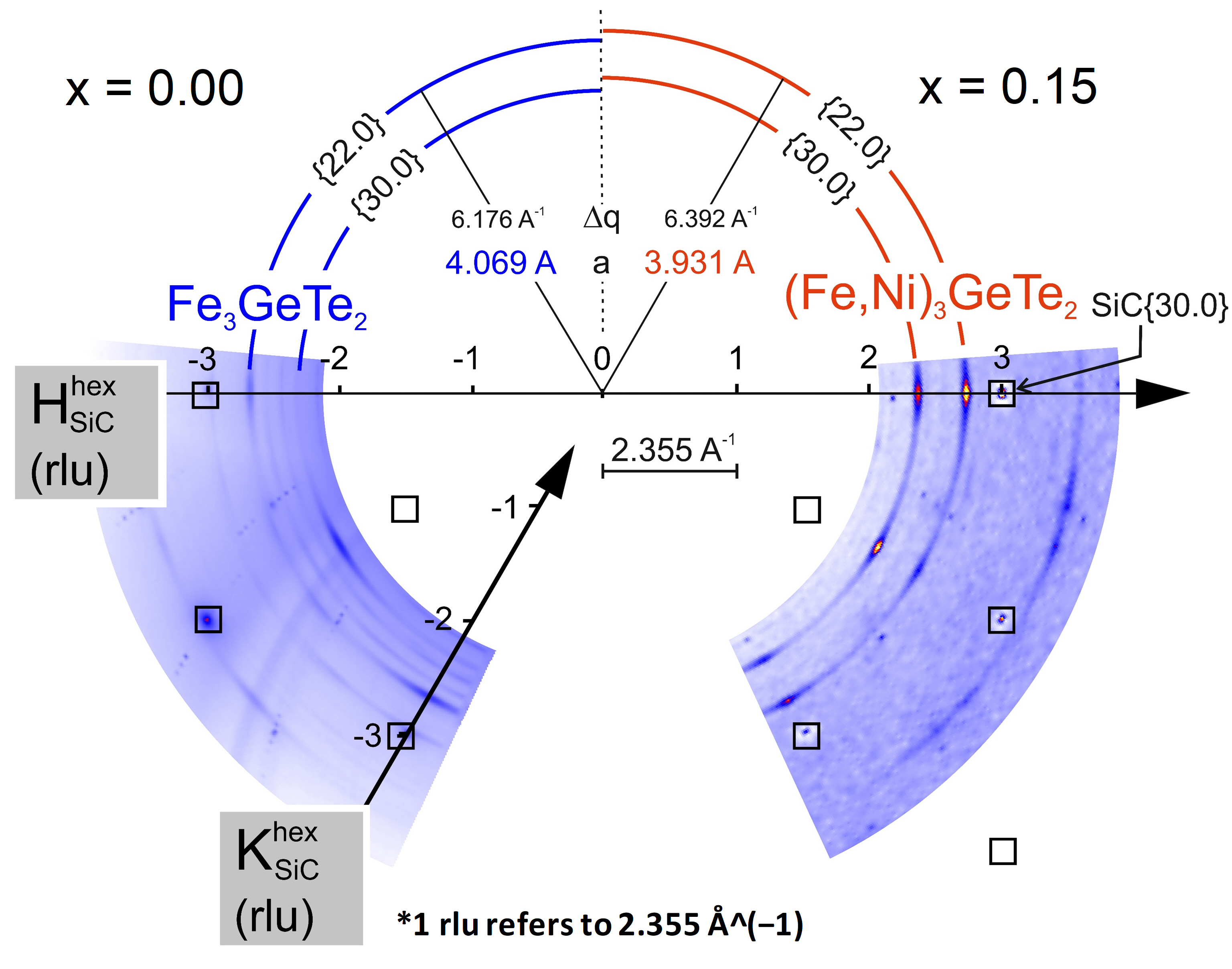}
\caption{\label{fig:GID} In-plane reciprocal space maps for two samples, showing a set of SiC substrate reflections (black squares) as references, and additional modulated arcs of intensity that can be attributed to epitaxial FGT and Ni-doped FGT (FNGT). In-plane lattice parameters are encoded by the arc diameter, while the azimuthal distribution yields the particular profile following the arc. Please note that the closely spaced multiple intensity maxima (especially in the left-hand figure) are artifacts.}
\end{figure*}
 Figure~\ref{fig:GID}, shows the diffusely scattered and color-coded intensity over a large area of reciprocal space as a function of H and K and at a vertical scattering vector L = 0 (i.e. in-plane). Anticipating at least six-fold symmetry, both diffraction patterns cover an azimuthal opening angle greater than 60°. Units are given in multiples of the reciprocal lattice units (rlu) of the underlying 4H-SiC substrate. Consequently, 1 rlu refers to 2.355 \AA$^{-1}$.  The diffraction patterns clearly reveal the fingerprints of both the pristine FGT and Ni-doped (x = 0.15) FGT (FNGT), as can be seen from the two arcs representing different net plane families. In-plane lattice parameters of 4.069 Å for FGT and 3.931 Å for FNGT were extracted from the \{22.0\} and \{30.0\} arcs for FGT (blue) and FNGT (red). This contraction of in-plane lattice parameter is consistent with the result obtained from RHEED streaks. Thus, we found that replacing Fe with Ni, in-plane lattice parameter of FGT unit cell decreases. Furthermore, based on the intensity profiles along the arcs, we can deduce the following preferential in-plane orientation of the layers: ${\rm FGT/FNGT}[11.0]~||~{\rm SiC}[10.0]$. Besides that, there is an additional, non-dominant type of domain azimuthal 30° off, thus ${\rm FGT/FNGT}[10.0]~||~{\rm SiC}[10.0]$. 

\section{Calculated lattice contraction with Ni doping.}
\begin{figure*}[htbp!]
\centering
\includegraphics*[width=0.9\linewidth]{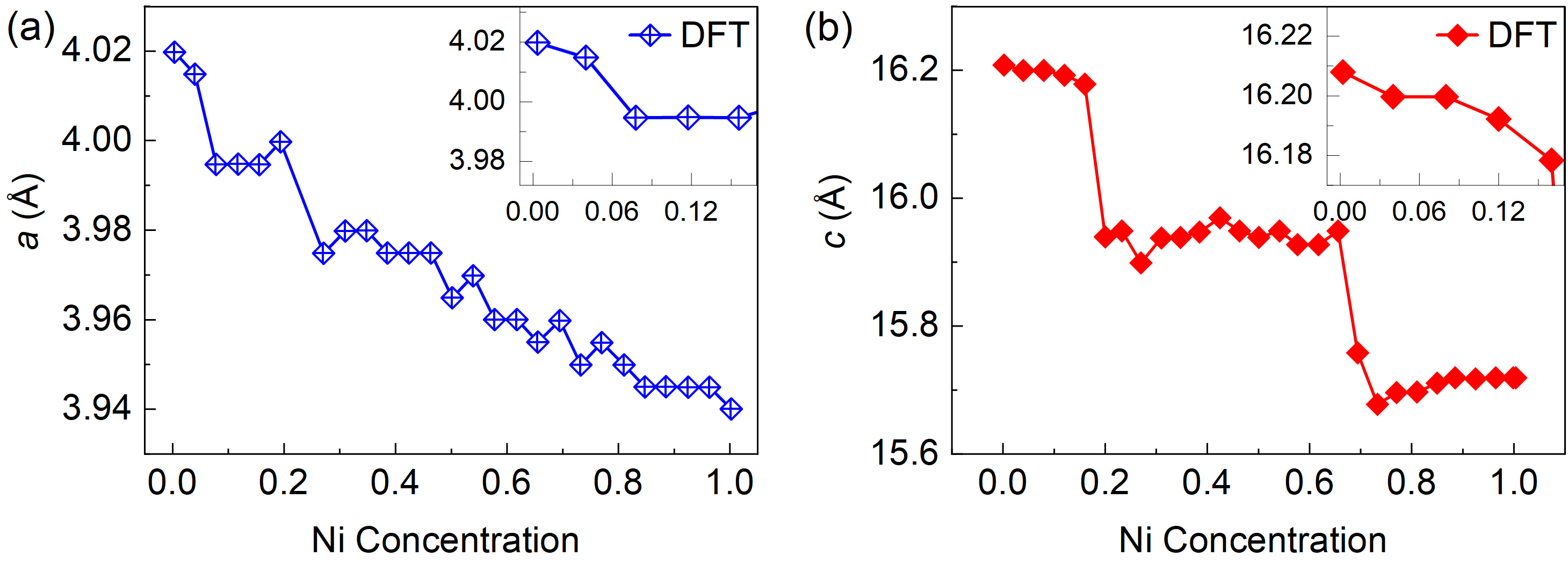}
\caption{\label{fig:S3} Variation of (a) in-plane and (b) out-of-plane lattice parameters as a function of Ni doping concentration in a pristine FGT structure, ranging from x = 0.0 to 1.0, where x = 1.0 corresponds to the NGT structure. The inset shows a zoomed-in view over the doping range used in the experiment.}
\end{figure*}
\begin{figure*} [htbp!]
\centering
\includegraphics*[width=0.99\linewidth]{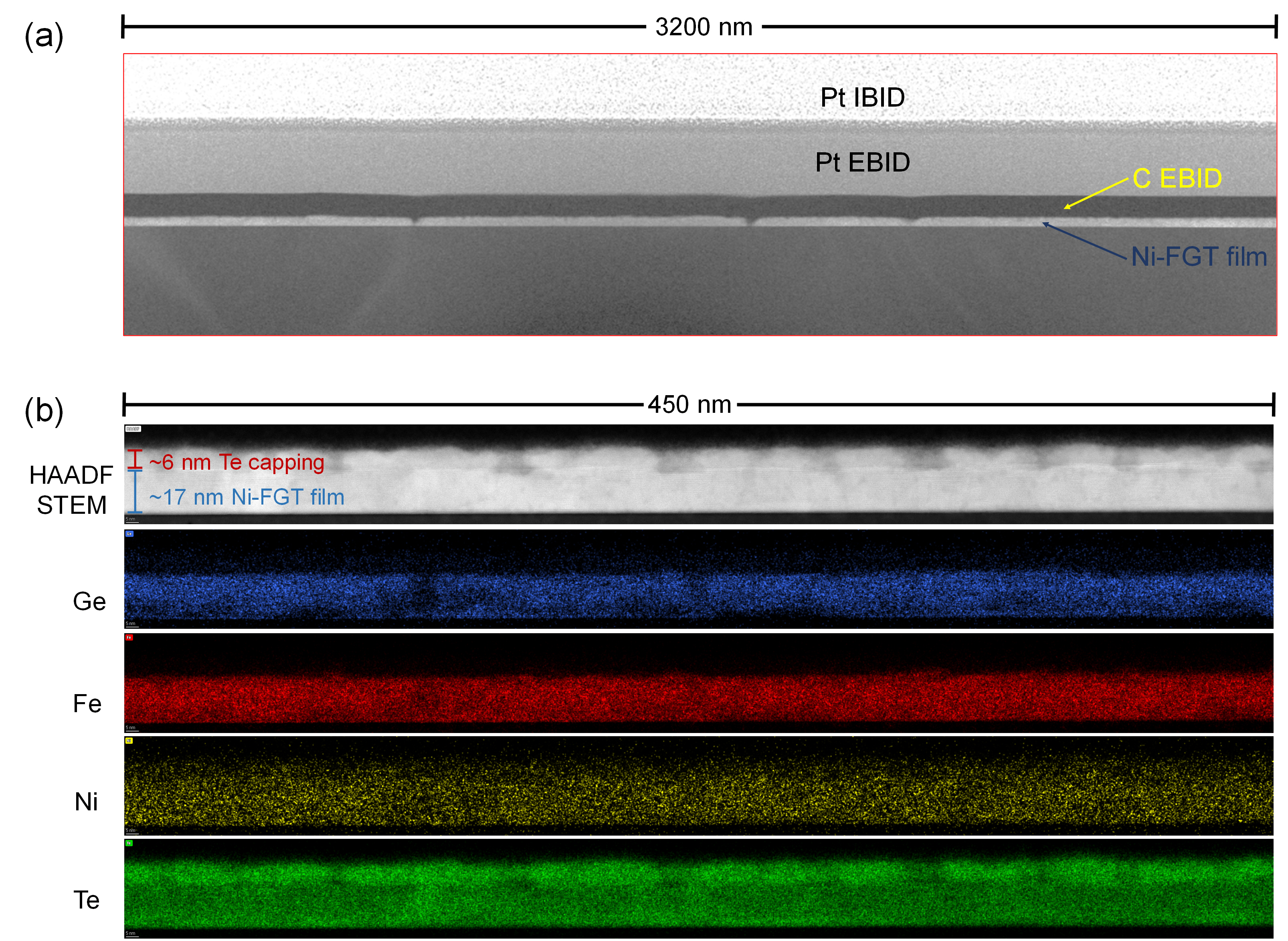}
\caption{\label{fig:S4} Cross-sectional TEM image of a sample prepared via dual-beam SEM/FIB (FEI Helios G4 CX), protected with EBID C/Pt and IBID Pt layers. (b) High-magnification TEM image of a Ni-doped FGT film (450 nm scale) and corresponding EDX elemental maps for Fe, Ge, Ni, and Te.}
\end{figure*}

Using DFT calculations, we studied the effect of Ni doping on structural modifications, upon $14.3\%$ Ni doping as shown in Figure~\ref{fig:S4}(a,b). To decouple the effect of Ni occupying sites, three scenarios are considered, where Ni sits in the vdW gap, Ni(1) or Ni(2) sites. Table~\ref{tab_lattice} shows the structural parameters found in these three scenarios. It was found that Ni doping in general shrinks the \textit{c} lattice parameter. However, the decrease upon occupation of Ni3 sites is more prominent. A comparison of these data with that of experiment given in Fig.~1(f) in main manuscript suggests an over estimation of \textit{c} in pristine FGT, twhich can be attributed to a fully stoichiometric simulation of pristine FGT, while in actual films imperfections are expected to exist, such as vacancies and stack faulting. Considering the $\sim0.06~{\text \AA}$ shrinkage along \textit{c}-axis upon $15\%$ experimental Ni-doped FGT, it can be inferred that a combination of Ni(1), Ni(2), and Ni(3) occupations should be expected in the experimental sample.

\begin{table}[htbp!]
\caption{Variation of in-plane and out-of-plane lattice parameter from pristine FGT to Ni-doped FGT films, by substituting Ni at different Fe sites.}
\centering 
\setlength{\tabcolsep}{52pt}
\renewcommand*{\arraystretch}{1.4}
\begin{tabular}{l|c|c}
\hline
\hline
Structure & \textit{a} & \textit{c} \\
 & (\AA) & (\AA) \\
\hline 
Pristine FGT        & 4.02 & 16.21  \\ 
14.3\% Ni(1) doped   & 4.00 & 16.18\\  
14.3\% Ni(2) doped   & 4.01 & 16.11  \\ 
14.3\% Ni(3) doped   & 4.04 & 15.86 \\ 
\hline
\end{tabular}
\label{tab_lattice}
\end{table}

\newpage
\section{FIB-Lamella and EDX mapping.}
\newpage
\section{Fe intercalation in pristine FGT}
\begin{figure*} [htbp!]
\centering
\includegraphics*[width=\linewidth]{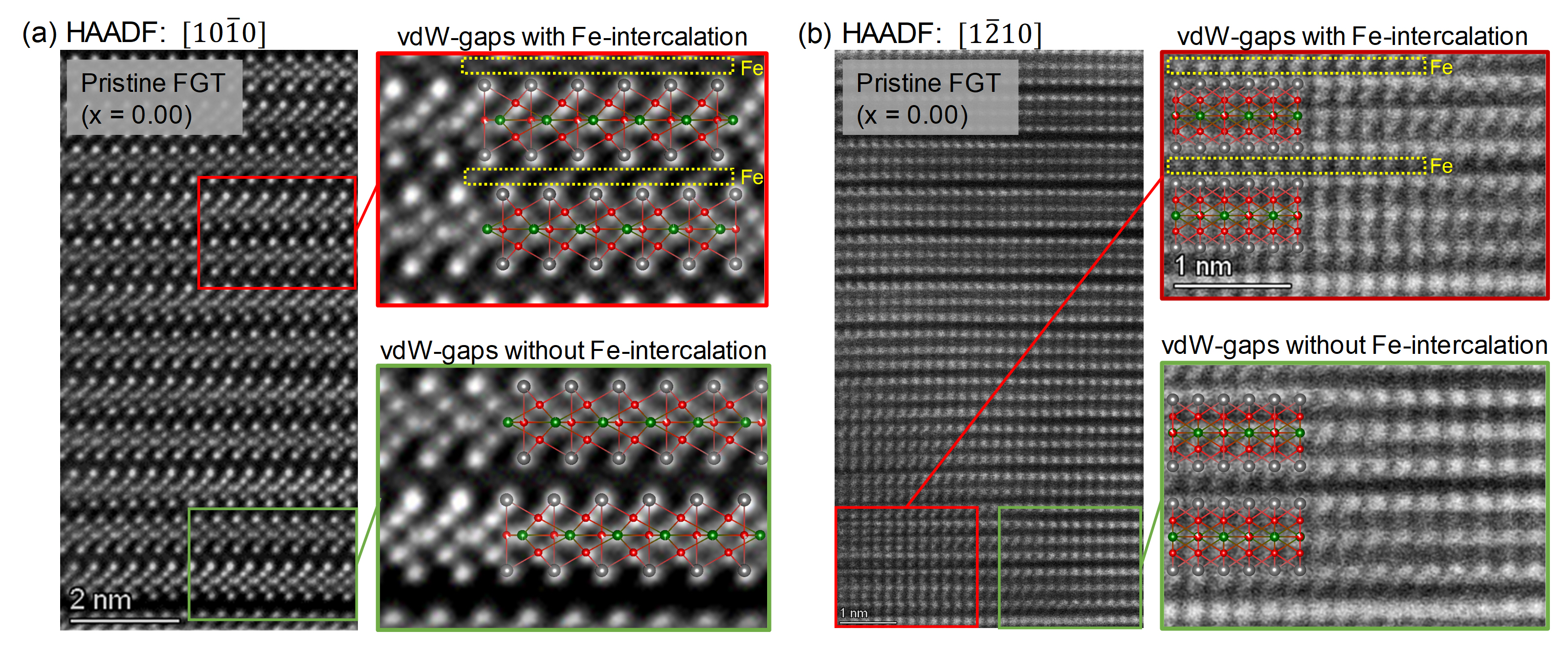}
\caption{\label{fig:STEM} Cross-sectional STEM-HAADF image for the pristine FGT along (a) [10$\bar{1}$0] and (b) [11$\bar{2}$0] crystallographic axes. The red (green) box shows the zoomed-in regions with (without) Fe intercalation in the vdW gaps.}
\end{figure*}
Figure~\ref{fig:STEM}(a) displays a high-resolution cross-sectional STEM-HAADF image of a pristine (x = 0.00) FGT film along both [10$\bar{1}$0] and [1$\bar{2}$10] crystallographic axes, confirming its atomically flat and layered hexagonal structure. Notably, in some localized regions [red box, Figure~\ref{fig:STEM}(b)], we observed excess Fe atoms occupying interstitial sites between Te-terminated layers, disrupting the vdW gaps in pristine FGT [see the yellow dotted in Figure~\ref{fig:STEM}(a, b)]. This Fe intercalation exhibits a spatially random distribution, which we attribute to growth kinetics and thermodynamic instabilities during synthesis. Such inhomogeneous intercalation introduce small local variations in magnetic exchange interactions, potentially explaining sample-dependent fluctuations in Curie temperature ($T_{\text{C}}$) and magnetic anisotropy in MBE-grown films~\cite{silinskas_self-intercalation_2024, wu2023fe}. These results show the critical role of vdW gap engineering via controlled Fe intercalation in tailoring the electronic and magnetic properties of FGT for spintronic applications.

\newpage
\section{Anomalous Hall effect (AHE)}
\begin{figure*} [b!]
\centering
\includegraphics*[width=0.7\linewidth]{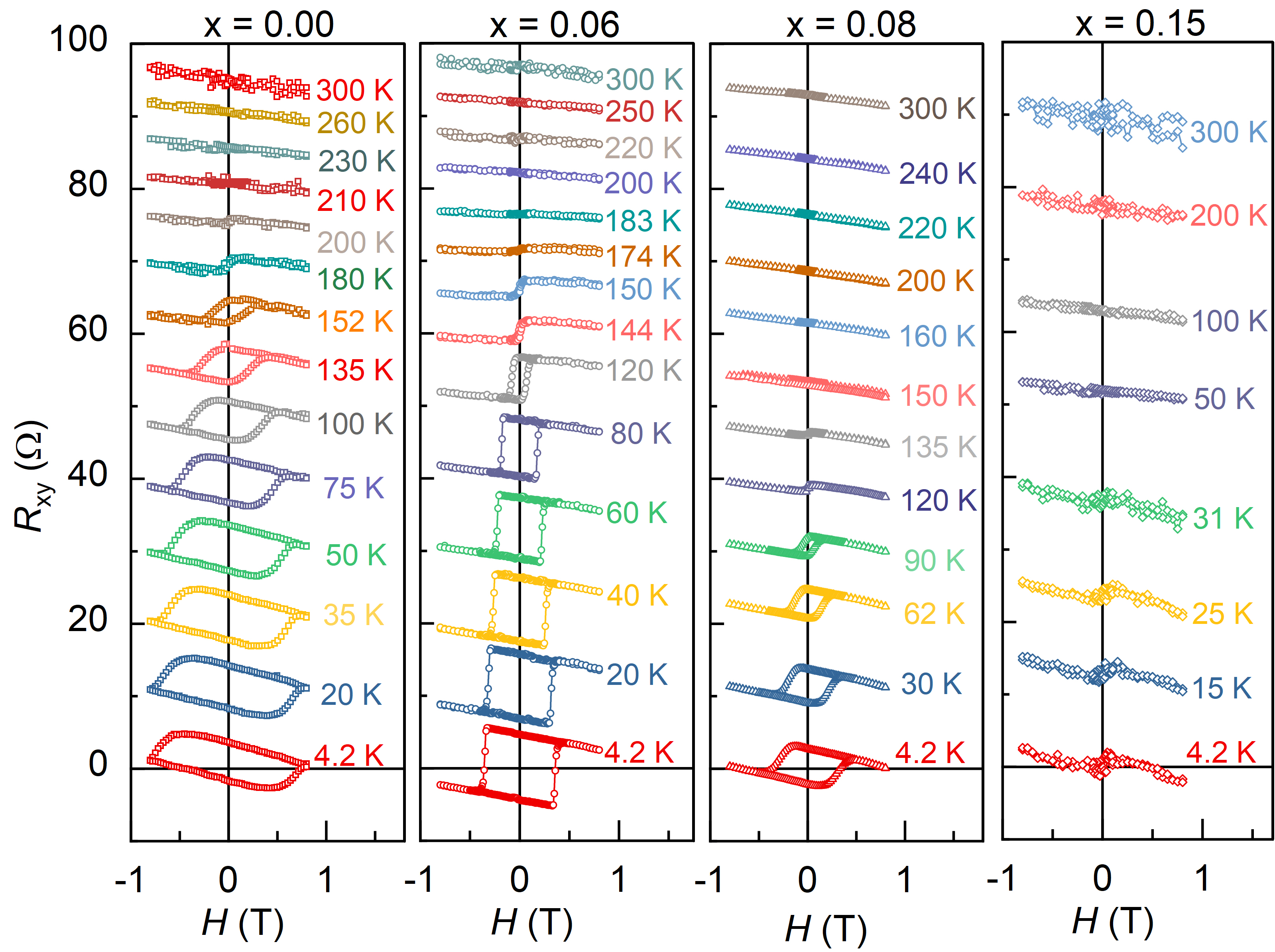}
\caption{\label{fig:AHE}The transverse Hall resistance $R_{\rm xy}$ versus magnetic field ($H$) measured at 4.2–300~K, for pristine FGT and three different Ni-doped FGT films. The plots are shifted vertically for clarity.}
\end{figure*}
AHE measurements were conducted across a temperature range of 4.2–300 K for the pristine (x = 0.00) Fe$_3$GeTe$_2$ and Ni-doped (x = 0.06 -- 0.15) [Fe$_{1-x}$Ni$_x$]$_3$GeTe$_2$ films (see Figure~\ref{fig:AHE}). The transverse Hall resistance $R_{\rm xy}$ was measured in Hall geometry, by applying a magnetic field perpendicular to the film and the charge current parallel to the surface of film, following the relation:~\cite{hurd_hall_1972};
\begin{equation}
 R_{xy} = R_0+ R_{\rm A},
\label{eqn:1}
\end{equation}
Here, the first term corresponds to the ordinary Hall resistance $R_{0}$, while the second term represents the anomalous Hall resistance $R_{\rm A}$. The values of $R_{\rm A}$ for all the films were extracted in the saturation magnetization regime of $R_{\rm xy}$ and plotted in  Figure 3(i-j) [main manuscript]. The value of carrier concentration ($n_e$) for all films was extracted using the parameter $R_{0}=1/n_{e}e$ ($e$ is the electronic charge), yielding values in the range $(1.5-6) \times 10^{14}/{\rm cm}^2$ across temperatures of 10–300~K. This $n_e$ represents the combined contribution from both the \textit{n}-type carriers in the graphene/SiC template and the FGT film grown on top. An independent measurement was carried out for bare graphene/SiC substrate and extracted a value of $n_e$ to be $\sim0.2\times 10^{14}/{\rm cm}^2$, approximately 5–30 times lower than that of the FGT/graphene stack. Therefore, a dominant observed negative slope in the $R_{\rm xy}$ plots in Figure~\ref{fig:AHE} thus arises from the parallel conduction channels of FGT and graphene/SiC, consistent with our earlier findings in similar graphene/FGT heterostructures~\cite{lopes2021large}.
\newpage
\section{AHE with different capping}
\begin{figure*} [htbp]
\centering
\includegraphics*[width=0.9\linewidth]{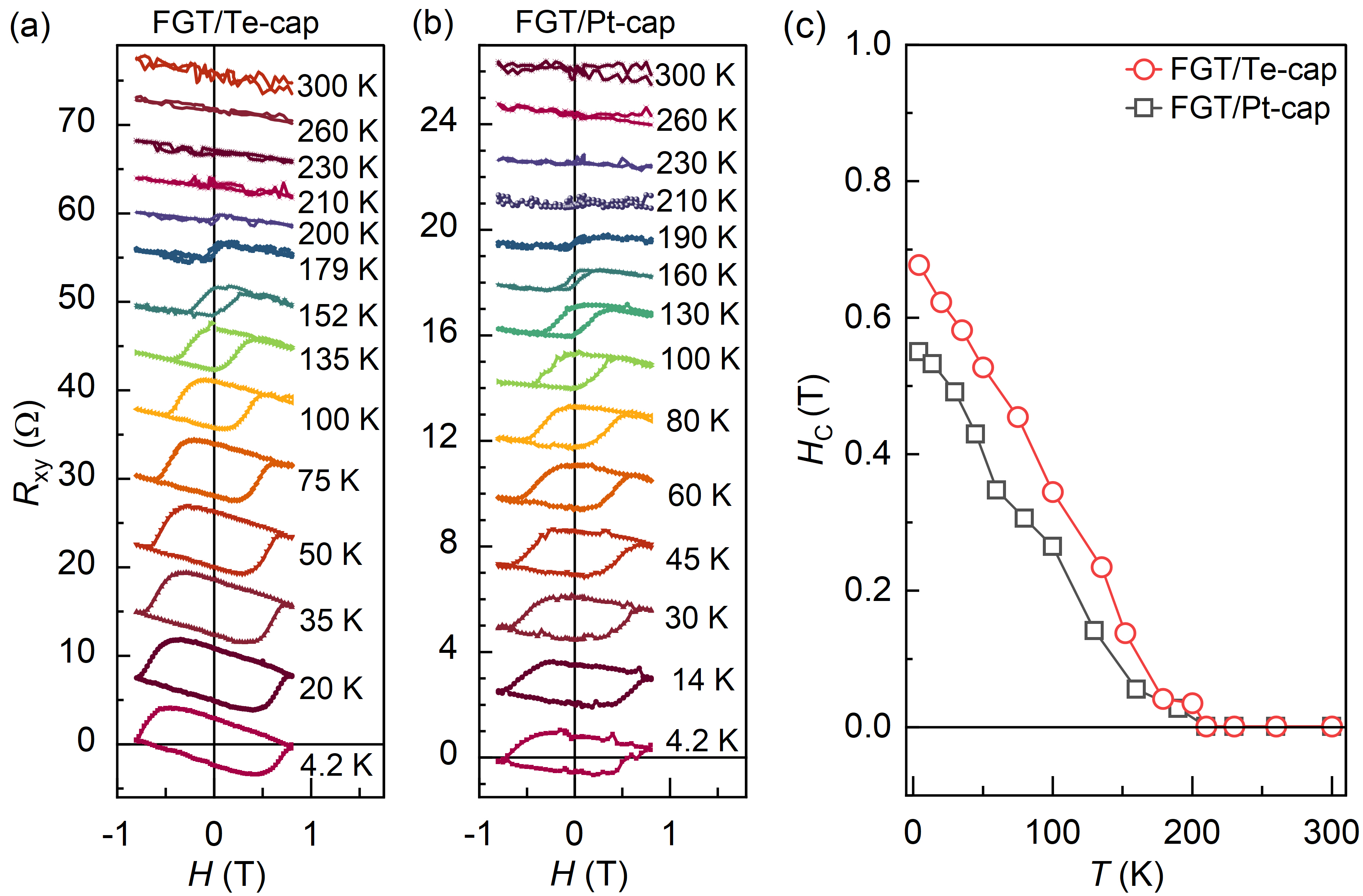}
\caption{\label{fig:FGT-Pt and FGT-Ta} Comparison of the transverse Hall resistance $R_{\rm xy}$ for the pristine FGT (x = 0.00) with (a) Te-capping and (b) Pt-capping.}
\end{figure*}

\newpage
\section{X-ray absorbtion spectra (XAS) at both Fe-and Ni-edge.}
\begin{figure*} [htbp]
\centering
\includegraphics*[width=0.9\linewidth]{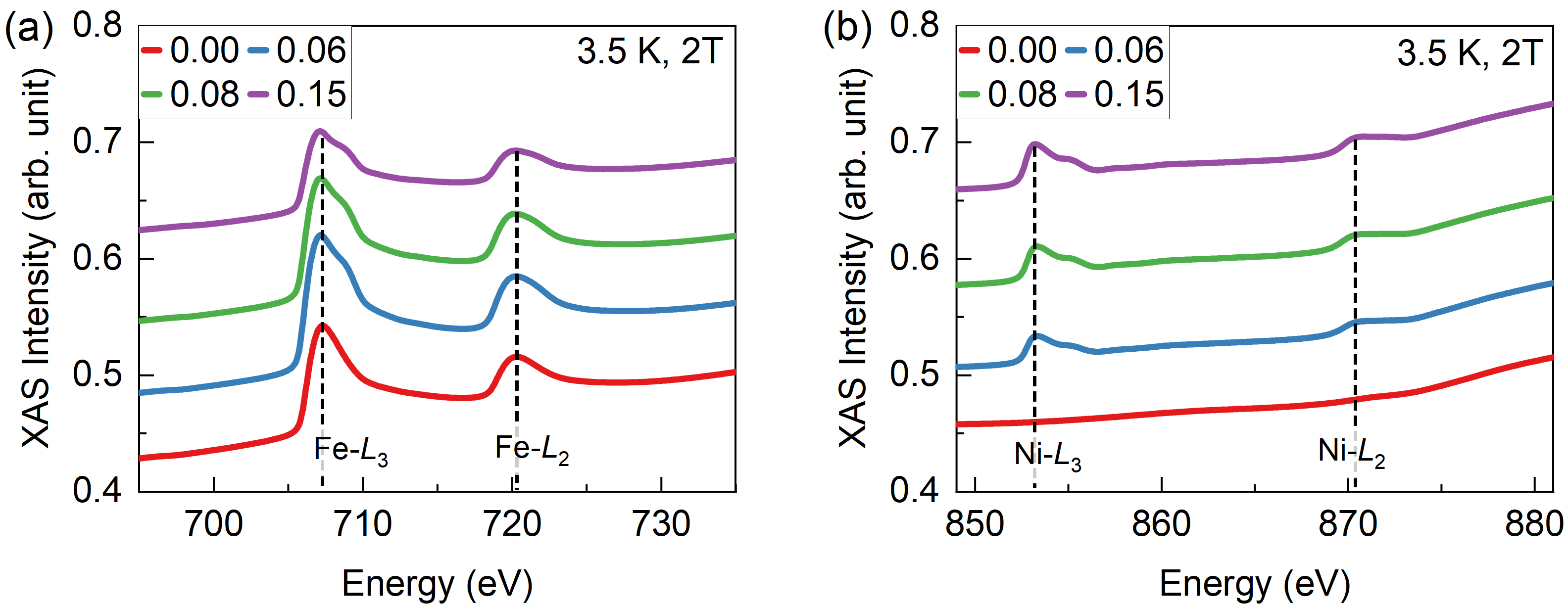}
\caption{\label{fig:S7} X-ray absorbtion spectra (XAS) for all the films measured at 3.5~K in the presence of magnetic field 2~T, at (a) Fe-$L_{\rm 2,3}$ and (b) Ni-$L_{\rm 2,3}$ absorption edge.}
\end{figure*}
\section{Dependence of $T_{\rm C}$ with Ni-doping.}
\begin{figure*} [htbp]
\centering
\includegraphics*[width=0.45\linewidth]{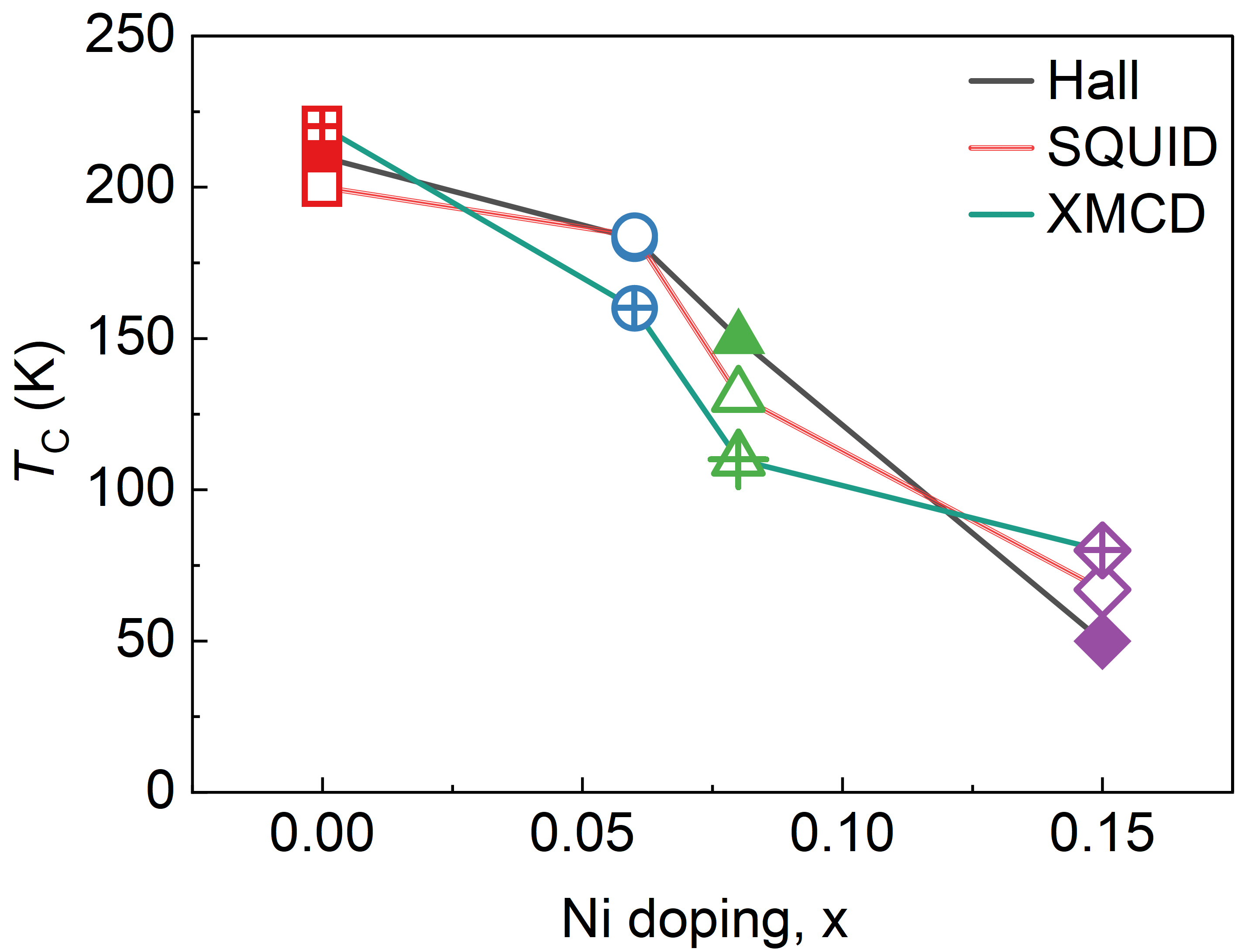}
\caption{\label{fig:S8} Dependence of $T_{\rm C}$ for pristine FGT and Ni-doping FGT films, obtained from different measurement techniques.}
\end{figure*}

\bibliography{reference.bib}

\end{document}